\documentclass[preprint,twocolumn]{rmaa}

\usepackage{paralist}

\usepackage{psfrag,color}

\usepackage[latin1]{inputenc}

\def\spose#1{\hbox to 0pt{#1\hss}}
\def\lta{\mathrel{\spose{\lower 3pt\hbox{$\mathchar"218$}}
     \raise 2.0pt\hbox{$\mathchar"13C$}}}
\def\gta{\mathrel{\spose{\lower 3pt\hbox{$\mathchar"218$}}
     \raise 2.0pt\hbox{$\mathchar"13E$}}}

\title{Effects of Magnetic Turbulence on the Dynamics of Pickup Ions in the 
Ionosheath of Mars}  

\author{H. Aceves,\altaffilmark{1}  M. Reyes-Ruiz \& C. E. Ch\'avez}

\altaffiltext{1}{Instituto de Astronom\'{\i}a, Universidad Nacional Aut\'o\-noma de M\'exico.}

\shortauthor{Aceves et al.}
\shorttitle{Magnetic turbulence and pickup ions in Mars}

\fulladdresses{
\item H\'ector Aceves, Mauricio Reyes, Carlos E. Ch\'avez: Instituto de Astronom\'\i a, Universidad Nacional Aut\'onoma de M\'exico, Apdo. Postal 106, Ensenada B.~C. 22800 M\'exico (\url{aceves,maurey,carlosepech@astrosen.unam.mx}). }

\listofauthors{H. Aceves, M. Reyes-Ruiz, \& C. E. Ch\'avez}
\indexauthor{Aceves, H.}
\indexauthor{Reyes-Ruiz, M.}
\indexauthor{Chavez, C. E.}

\abstract{
We study some of the effects that magnetic turbulent fluctuations have on the dynamics of pickup $O^+$ ions in the magnetic polar regions  of the Mars ionosheath. In particular we study their effect on the bulk velocity profiles of ions as a function  of altitude over the magnetic poles, in order to compare them with recent Mars Express data; that indicate that their average velocity is very low and essentially in the anti-sunward direction. We find 
that, while magnetic field fluctuations do give rise to 
deviations from simple ${\bf E} \times {\bf B}$-drift gyromotion,
even fluctuation amplitudes much greater than those of {\it in 
situ} measurements are {\it not} able to reproduce the vertical velocity profile of  $O^+$ ions. We conclude that other physical mechanisms, different from a pure charged particle dynamics, are acting on pickup ions at the Martian terminator. A possibility being a viscous-like interaction between the Solar Wind and the Martian ionosphere at low altitudes.
}

\resumen{Se estudian los efectos que fluctuaciones magn\'eticas turbulentas tienen sobre la dinámica de iones {\it pickup} de $O^+$ en las regiones polares magnéticas de la ionofunda de Marte. En particular estudiamos su efecto
 en los perfiles de velocidad promedio como funci\'on de la altitud sobre los polos, con el fin de comparar con datos reciente del Mars Express;  que indican que su velocidad promedio es peque\~na y esencialmente en la direcci\'on antisolar. Se encuentra que, aunque las fluctuaciones magn\'eticas resultan en desviaciones del simple movimiento de arrastre girotr\'opico, aun amplitudes de fluctuaci\'on m\'as grandes que las medidas, {\it no} son capaces de reproducir los perfiles verticales de velocidad de los iones de  $O^+$. Conclu\'imos que otro mecanismo f\'{\i}sico, diferente a uno puramente de din\'amica de part\'{\i}culas cargadas, est\'a actuando sobre estos iones en el terminador Marciano. Una posibilidad es que exista una interacci\'on tipo-viscosa entre el viento solar y la ionosfera Marciana a bajas altitudes.
}

\addkeyword{Physical Processes: turbulence, magnetic fields, acceleration of particles}
\addkeyword{Solar System: planets: Mars}
\addkeyword{Methods: numerical}

\begin{document}
\maketitle

\section{Introduction}\label{sec:intro}

The interaction of the solar wind (SW) with Solar System objects with a negligle intrinsic magnetic field and an atmosphere (e.g.~Venus and Mars) is currently being investigated by several space missions (e.g.~Venus and Mars Express), that will enhance our  understanding of plasma interaction processes in such environments (e.g. Ma et~al.~2008). In particular, a comparison of different models for the global interaction of the solar wind (SW), a weakly collisional plasma (e.g. Marsh~1994, Echim et al.~2011), with Mars has been done by Brain et al.~(2010) that included MHD, multi-fluid and hybrid models. All these models, as well as the gasdynamic convected magnetic field models (e.g. Belotserkovskii et al.~1987, Spreiter \& Stahara~1994),  have greatly  advanced  our knowledge of the complex interaction of the SW plasma and the Martian ionosphere.

An important component of the Martian plasma environment, specially at low and mid altitudes, is the population of oxygen ions ($O^+$) produced by the UV solar radiation and charge exchange reactions. These are picked up by the SW convective electric field and removed from the planet,  in turn affecting the solar wind interaction with Mars (e.g. Ledvina~et~al.~2008, Withers~2009)

\begin{figure}[!t]
\centering
\includegraphics[width=\columnwidth]{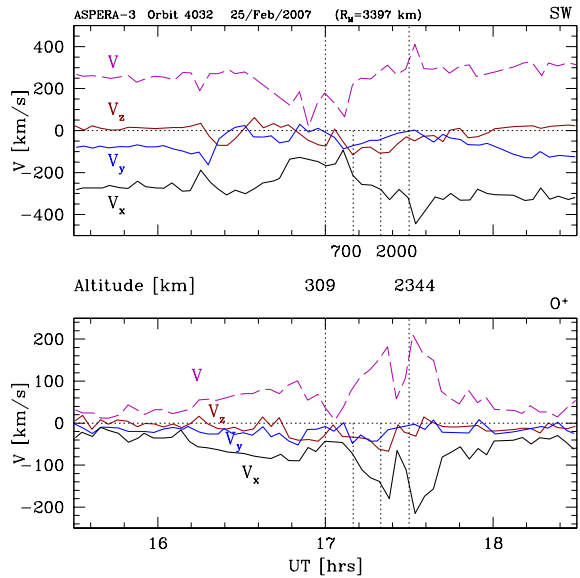}
\caption{ASPERA-3 {\it in situ} measurements of the velocities profiles of SW protons ({\it top}) and $O^+$ pickup ions ({\it bottom}) over the terminator of Mars;  coordinates are in a MSE system, and the data is taken from PdT09. Four altitudes of the spacecraft are indicated as dotted vertical lines. The dominance of the $x$-component of the velocity of $O^+$ ions  over its $z$ component is very clear, contrary to what is expected from a pure $\mathbf{U}\times \mathbf{B}$--drift.} 
\label{fig:aspera}
\end{figure}

Recently, Perez-de-Tejada et al.~(2009; PdT09) have pointed out that the analysis  of Mars-Express measurements, conducted with the ASPERA-3 instrument (Barabash et al.~2006), indicates that the bulk velocity of $O^+$ ions, presumably of ionospheric origin, as they stream at low altitude over the magnetic poles of the planet is mostly in the anti-sunward direction (see Figure~\ref{fig:aspera}). 
This result is unexpected in the context of a SW-ionosphere interaction in which the acceleration of pickup ions is essentially due to the action of the convective electric field, $\mathbf{E}$. 
As shown in Reyes-Ruiz~et~al.~[2010b (RAP10)], in the context of a simplified  model for the geometry of the IMF and SW flow in the ionosheath of Mars, the bulk velocity of pickup $O^+$ ions in gyromotion, as they pass over the magnetic poles of the planet, is strongly dominated by the vertical component ($v_z$, normal to the plane of the ecliptic). 

A qualitatively similar conclusion to the above can also be reached from the results of MHD simulations using more complicated IMF and Martian magnetic field models to study the dynamics of heavy ions in the region (e.g. Fang~et~al.~2008), and seen in earlier $O^+$ pickup ion trajectories in a draped magnetic field model (e.g. Luhmann \& Schwingenschuh~1990, Luhmann~1990, Lichtenegger~et~al.~2000). In RAP10 it is argued that a largely horizontal bulk velocity of $O^+$ ions can be  
explained in terms of a viscous-like interaction between the SW and ionospheric plasmas, as found in the numerical simulations of 
Reyes-Ruiz et al.~(2009).

It has been suggested that magnetic field fluctuations, as those known to 
be widespread in the ionosheath of Mars and other non-magnetic bodies 
(Tsurutani~et~al.~1995, Nagy~et~al.~2004, Grebowsky et~al.~2004, Espley~et~al.~2004, V{\"o}r{\"o}s et~al.~2008), may be responsible for modifying the gyromotion 
of pickup ions, enhancing pitch-angle scattering and ion heating and 
acceleration (e.g. Wang~et~al.~2006 and references therein). 
 Najib~et~al.~(2011) have noted more recently that the presence of wave activity and turbulence lead to wave-particle interactions that resemble collisions, hence probably modifiyng ion distribution function. In this paper we study the possibility that such modifications of simple gyromotion due to turbulence-like fluctuations of the magnetic field, in the ionosheath over the magnetic poles of Mars, lead to a bulk velocity profile for $O^+$ ions as that reported by PdT09 from the ASPERA-3 measurements.

 It is worth emphasizing that global MHD or Hybrid simulations analyzing the motion of ionospheric ions (Nagy et~al.~2004, Fang et~al.~2008, Brain~et~al.~2010, Kallio~et~al.~2010, Najib~et~al.~2011, and the review by Nagy et~al.~2004,) do not include the effect of small-scale turbulent fluctuations of the magnetic field, known to exist in the ionosheath of nonmagnetic Solar System bodies. These models essentially use the $\mathbf{B}$ field obtained directly from MHD simulations to follow the dynamics of the $O^+$ ions. Hence, our approach is not physically modeling the same phenomena that the above models are able to investigate. On other hand, do not consider the effect the crustal field (e.g.~Acu\~na~et~al.~1998, Zhang \& Li~2009) can have on the ion dynamics.

This paper is organized as follows. In $\S$2 we present the basic equations
and methodology used for our analysis, including the prescription for constructing the turbulent magnetic fields employed here. Our main results are
presented in $\S$3 for several cases having different properties of the power spectrum of magnetic fluctuations, and a comparison with measurements is done.  Finally, our concluding remarks and conclusions are presented in $\S$4.

\section{Model}

\begin{figure}
\centering
\includegraphics[width=8cm]{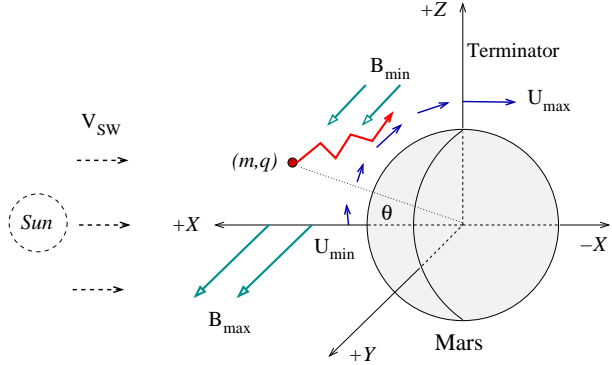}
\caption{Configuration used to study the motion of pickup ions, of mass $m$ and charge $q$. The $X$-axis  points toward the Sun, the $Z$-axis points perpendicular to the ecliptic and the $Y$-axis completes a right-hand triad. The draped magnetic field around Mars goes from a minimum to a maximum at the equator and pole, respectively. The SW velocity rises from a minimum to a maximum at the equator and pole, respectively. The angle $\theta$ is the solar zenith angle (SZA). The radius of Mars is taken to be $R_M=3397\,$km. Only the ``northern'' hemisphere is considered in this work.}
\label{fig:setup}
\end{figure}

We describe here the model used to study the motion of charged
particles ($O^+$ pickup ions) in a stationary background magnetic field  with a fluctuating magnetic and electric field components. The magnetic field configuration and the SW flow properties are taken to represent, in a first approximation, conditions present in the ionosheath of Mars.
 The geometry of our model is shown schematically in Figure~\ref{fig:setup}, with several quantities identified below.

\subsection{Equation of Motion}

We follow the three-dimensional motion of particles in a time-dependent
 fluctuating magnetic field and its corresponding induced electric
 fluctuation. We will refer to such fluctuations as ``turbulence'', meaning by this that it has a particular power spectrum density distribution as discussed below. We 
 treat only the non-relativistic motion of the pickup ions. We consider
 that the dynamics of these ions is determined solely by an  
 electromagnetic interaction.

The equation of motion for a particle of mass $m$
and charge $q$, in a region with electric $\mathbf{E}$ and magnetic $\mathbf{B}$ fields is given, in the non-relativistic limit, by the Lorentz equation (e.g. Jackson 1975):
\begin{equation}\label{eq:motion}
\frac{d }{d t} (m  \mathbf{v}) = q \left( \mathbf{E} +
\frac{\mathbf{v}}{c} \times \mathbf{B} \right) \,.
\end{equation}
For our problem at hand, the
magnetic field is assumed to be of the form: 
\begin{equation}
\mathbf{B}(\mathbf{r},t)= \mathbf{B}_0 + \delta
\mathbf{B}(\mathbf{r},t) \,,
\label{eq:totBfield}
\end{equation}
where $\mathbf{B}_0\!=\!\mathbf{B}_0 (\mathbf{r}) $ is non-homogeneous stationary background field, while $\delta \mathbf{B}(\mathbf{r},t)$ represents the turbulent part; with a zero-mean average value. No background
electric field is considered here, so only that $\mathbf{E}$ resulting from the
fluctuating $\delta \mathbf{B}$ will be taken into account in the equation of motion (see below).

\subsection{Steady Background Fields}

The steady fields determining the dynamics of the pickup ions are a
non-homogeneous background magnetic field $\mathbf{B}_0$ and the streaming SW velocity $\mathbf{U}_0$; see Figure~\ref{fig:setup}.

We are interested mainly in the dynamics of the ions over the terminator, hence restrict our simulation region to a radial interval and $y$-coordinates given by, respectively: 
$$
r \in [R_p,R_{out}], \quad \mathrm{and} \quad
y\in [-\frac{R_p}{3},+\frac{R_p}{3}]\,,
$$
where $r^2=x^2 +z^2$,  $R_p$ is the radius of the planet and $R_{out}$
is an external boundary; to be defined below. The regions $x\!>\! 0$ and $z\!>\! 0$ are the only ones considered for our numerical integrations.

The non-homogeneous background magnetic field used here resembles that
present in non-magnetic planets.  The magnetic field $\mathbf{B}_0$ in the
$Y$-direction is taken to be essentially constant, but with a
dependence on the coordinates $(x,z)$. For simplicity, we
adopt the following analytical expression (RAP10) for this magnetic field, in component form:
\begin{equation}
\mathbf{B}_0 =
\left[
\begin{array}{c}
0  \\
(B_{max}-B_{min}) \cos \theta + B_{min} \\
0 \\
\end{array}
\right] ,
\label{eq:B0field}
\end{equation}
where $\cos \theta=x/r$, and $\{B_{min},B_{max}\}$ correspond to
values of the magnetic field a the polar $(\theta=\pi/2$) and
sub-solar points ($\theta=0$), respectively; see Figure~\ref{fig:setup}.

The velocity field $\mathbf{U}_0$ of the SW plasma around the planet is taken
to increase sinusoidally in magnitude from the equator to the pole
as:     
\begin{equation}
\!\! \mathbf{U}_0 \! =\!
\left[
\begin{array}{c}
- [ (U_{max}-U_{min}) \sin \theta + U_{min} ]    \sin \theta  \\
0 \\
+ [  (U_{max}-U_{min}) \sin \theta + U_{min}   ]  \cos \theta \\
\end{array}
\right] ,
\label{eq:Ufield}
\end{equation}
where  $\{U_{min},U_{max}\}$ correspond to
values of a sub-solar points ($\theta=0$) and polar  $(\theta=\pi/2$)
points, respectively, and $\sin \theta = z/r$.

\subsection{Turbulent Magnetic Field}

There is at the time being no general theory of turbulence in plasmas and it appears to be no unique way to describe it (e.g.~Borovsky \& Funsten~2003, Cho et~al.~2003, Zhou et~al.~2004, Horbury et~al.~2005,  Galtier~2009).  On other hand, for example, the  value of spectral index of the SW turbulence being of a Kolmogorov or Iroshnikov-Kraichnan type or its nature is still a matter of investigation (e.g. Bruno \& Carbone~2005, Ng~et.~al.~2010).

In the case of homogeneous turbulence one may construct the  magnetic field  fluctuation   $\delta \mathbf{B}$ at a particular spatial point  by adding the contribution of a large set ($N_m$) of plane waves with different wave-number
($k$). This can be done, for example, by means of a Fourier transform on $P(k)$  (e.g. Owens~1978) or by orienting randomly in space the wave-vector
number (e.g. Giacalone \& Jokippi 1994), with  the field 
amplitude satisfying a particular form of the power spectrum.

Both approaches above have
their pros and cons (e.g.~Casse~et~al.~2002) and have been used in different
works on the effects of turbulence. For example, in cosmic
ray dynamics  in the radio lobes of galaxies (e.g.~Fraschetti \&
Melia~2008, O'Sullivan et al.~2009) or the acceleration of particles
in shocks (e.g. Muranushi \& Inutsuka~2009). Here we consider the propagation of one-dimensional turbulent plane waves traveling in each perpendicular direction in turn, and assume homogeneous turbulence. This will allow us to assess the effect of maximum turbulence effects along each direction on the dynamics of the $O^+$ ions; which is the main objective of this work.

 We write the  magnetic fluctuation field at a particular point in space as a
 superposition of transverse plane waves propagating, for example, in the
  $\pm Z$--direction.  
 Hence we  write a $Z$--propagating wave  with arbitrary polarization
 (e.g.~Jackson~1975, Damask~2005)   as the real part of: 
\begin{equation}\label{eq:planewave}
\delta \mathbf{B} = 
\left[
\begin{array}{c}
\delta B_x  \\
\delta B_y  \\
\end{array}
\right]
=
\sum_{j}^{N_m}  \,  \delta b_j \,  |p\rangle \, {\rm e}^{{\rm i} (k_j z -
  \omega_j t)} \,, 
\end{equation}
where the polarization state or Jones vector $|p\rangle$ in the $XY$--plane is
given by
\begin{equation}\label{eq:phases}
 |p\rangle \equiv 
\left[
\begin{array}{c}
 p_x  \\
 p_y  \\
\end{array}
\right] = 
\left[
\begin{array}{c}
 \cos \varphi \; {\rm e}^{{\rm i} \phi_x}  \\
\sin \varphi \; {\rm e}^{{\rm i} \phi_y}   \\
\end{array}
\right]\,;
\end{equation}
with the $\varphi$'s and $\phi$'s being parameters describing the
polarization state of the wave. 
For example, for $\phi_x=\phi_y$ we
have a linearly polarized wave inclined an angle $\varphi$, while if
$\phi_x$ and $\phi_y$ differ by $\pi/2$ circular polarization is 
produced; otherwise elliptical polarization is achieved.

\begin{figure*}[!th]
\centering
\includegraphics[width=0.49\textwidth]{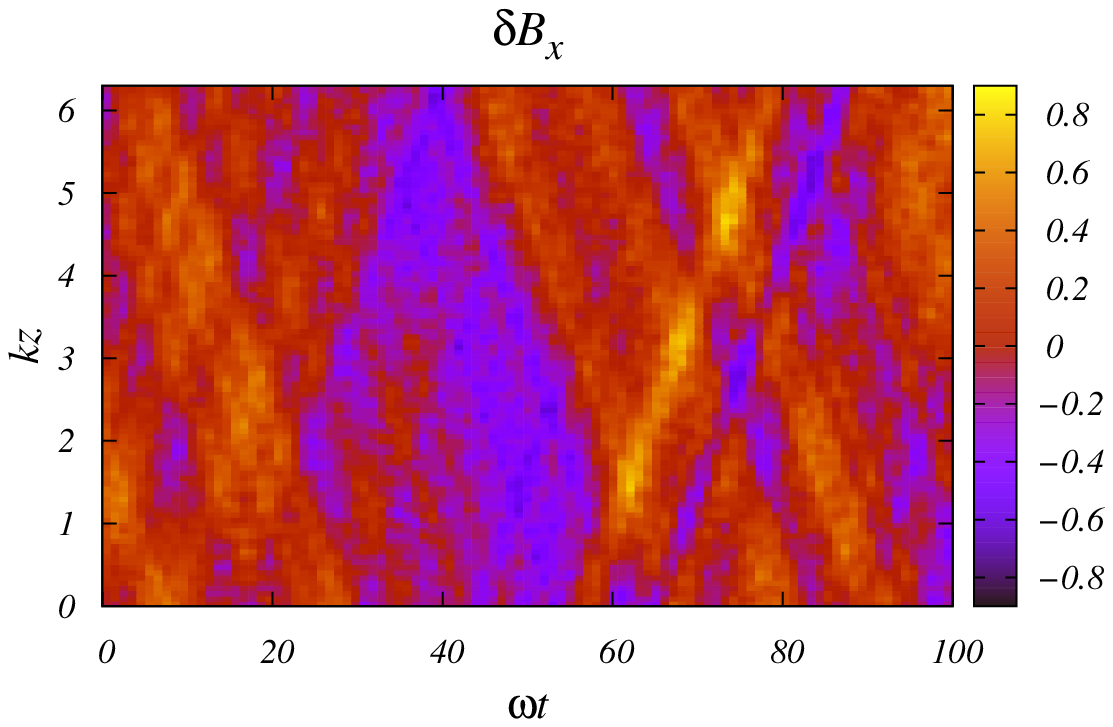}
\includegraphics[width=0.49\textwidth]{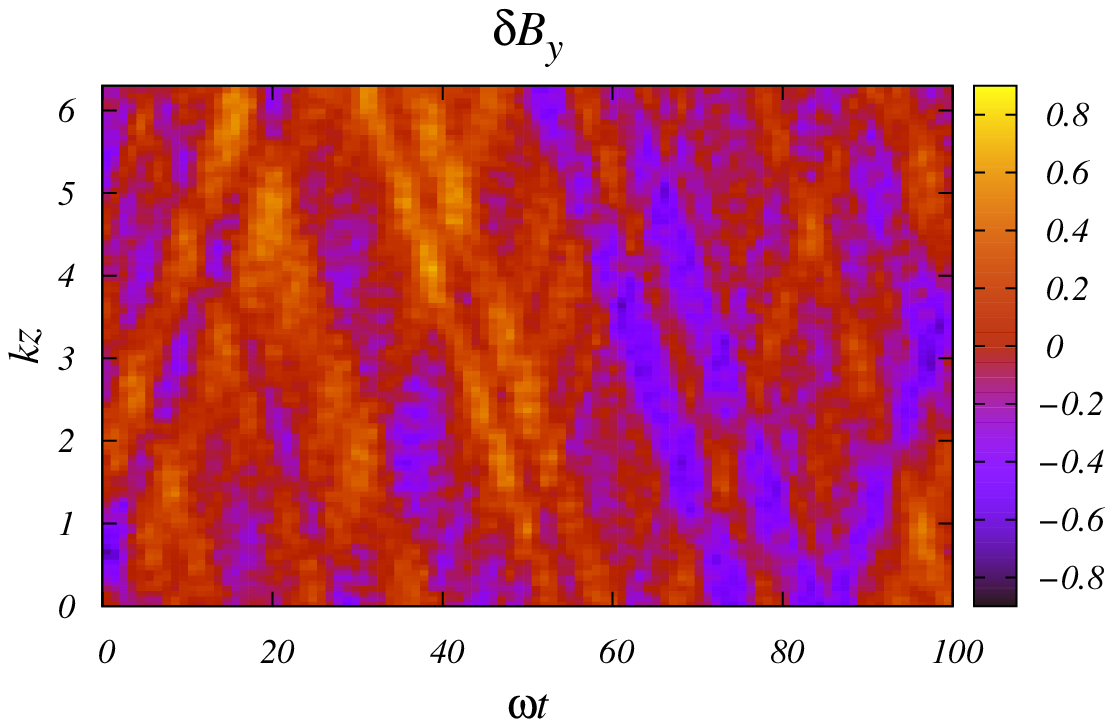}
\vspace{-18pt}
\caption{Intensity plot of a particular turbulent magnetic field realization of plane waves traveling in the $Z$--direction at different times. The $X$--component is shown at the left figure while the $Y$-component at the right figure. The total field variance here is $\langle  \delta B^2\rangle=1$ and $\gamma=2$. Units are arbitrary.} 
\label{fig:campo}
\end{figure*}

The real value functions $\delta b_j = \delta b(k_j)$ in
(\ref{eq:planewave})  are the fluctuating field amplitude 
corresponding to the $j$-th mode of the $m$-number of modes $N_m$. The frequency of oscillation of each
mode is $\omega_j=\pm v_{ph} k_j$, with $v_{ph}$ being the
phase-velocity of the $j$-th mode.  For definitiveness, the phase-velocity of the perturbations is taken here to be the
Alfv\'en velocity of the plasma (i.e., $v_{ph}=v_A$). The sign in the
expression for $\omega_j$ accounts for waves traveling both in the
positive or negative $Z$--direction, respectively. 
The form of turbulence represented in (\ref{eq:planewave}) satisfies,
by construction, the divergence-free character of the 
magnetic field fluctuation ($\nabla \cdot \delta
{\mathbf B}=0$).  A similar construction is done for each orthogonal propagation direction considered here.

The amplitudes $\delta b_j$ in (\ref{eq:planewave})  are 
chosen to satisfy a particular one-dimensional power spectrum of fluctuations $P(k)$.  We assume here that the power spectrum density $P(k)$ is given by a power-law: 
\begin{equation}\label{eq:spectrum}
P(k) = P_{N} \, \langle \delta B^2\rangle \,  k^{-\gamma} \;,
\end{equation}
where $\langle  \delta B^2\rangle$ is the magnetic field fluctuation variance,
$\gamma$ the spectral index and $P_{N}$ a normalizing constant.  For $\gamma=5/3$ we have a Kolmogorov (1941) turbulence and for
$\gamma=3/2$ the Iroshnikov-Kraichnan type spectrum (Iroshnikov 1963,
Kraichnan 1965). The normalization of the spectrum is 
such that
\begin{equation}\label{eq:spectrum2}
\int_0^\infty  P(k) \,  {\rm d} k =  \langle \delta B^2\rangle \,.
\end{equation}
In practice, the limits
of integration in (\ref{eq:spectrum2}) go from a $k_{min}$ to a $k_{max}$
dictated essentially by numerical criteria of the calculations.

The discrete amplitudes $\delta b_j$ in (\ref{eq:planewave}) can now be
obtained from the power spectrum (\ref{eq:spectrum}). Using the
orthogonality of the polarization state vector,
$$
\langle p | p \rangle \equiv [ p_x^* , p_y^* ] 
\left[ 
\begin{array}{c}
 p_x  \\
 p_y  \\
\end{array}
\right] = 1 \,,
$$
where $\langle p |$ is the complex conjugate (dual) vector of $| p
\rangle$, and $\langle \delta B^2\rangle = \delta \mathbf{B} \cdot 
\delta \mathbf{B}^\dag$, and assuming statistical independence among the
different wave modes, we have 
\begin{equation}\label{eq:amplitudes}
\sum_{j=1}^{N_m} P(k_j) \Delta k_j  = \sum_{j=1}^{N_m} \delta b_j^2  
\;\; \to \;\;
 \delta b_j^2  =  P(k_j) \Delta k_j \,.
\end{equation}

The numerical realization of our turbulent magnetic field is obtained as
follows.  Given a mean fluctuation $\langle \delta B^2\rangle$,  a spectral index $\gamma$ and $\{k_{min},k_{max}\}$ for the
fluctuations, a normalization constant $P_N$ is determined from  (\ref{eq:spectrum2}). We choose a
set of $N_m$ equally spaced wave-numbers in  $\log k$--space in the relevant interval. From this set of $k$ values, the amplitudes
$\delta b_j$ are obtained using (\ref{eq:amplitudes}). Since we are not interested in any particular effect of the polarization state of the turbulent waves, we choose  random phase values  $\phi_x$ and $\phi_z$
and angle $\varphi$  in (\ref{eq:phases}) 
 uniformly distributed in $[0,2\pi]$ for each $k_j$ value.

At each point, for example, $z$ and time $t$ the sum in
(\ref{eq:planewave}) is evaluated, and
the $\delta B_x$ and $\delta B_z$ components are obtained. In this form,
the total magnetic fluctuation $\delta \mathbf{B}$ is
constructed; which is to be added to the inhomogeneous background
field $\mathbf{B}_0$ of equation (\ref{eq:totBfield}).  A similar construction is done for each propagation direction of turbulent waves when required. 
In Figure~\ref{fig:campo} we show a particular random 
realization of a trave;omg turbulent field.

\subsection{Electric Field}

In the stationary reference frame the electric field depends on the local total velocity field of the SW plasma $\mathbf{U}$ around the planet
 and the local magnetic field. In the non-relativistic limit this leads to
\begin{equation}
 \mathbf{E}(\mathbf{r},t) = -\frac{1}{c} \mathbf{U} \times
 \mathbf{B}(\mathbf{r},t) \,,
\label{eq:efield}
\end{equation}
where $\mathbf{B}$ is the total magnetic field obtained from equations
(\ref{eq:planewave}) and (\ref{eq:B0field}). The total velocity field $\mathbf{U}$ is obtained as $\mathbf{U}= \mathbf{U}_0 + \delta \mathbf{U}$, were we consider that $\delta \mathbf{U} (\mathbf{r},t) = -v_A \delta \mathbf{B}/B_0$ as in a MHD Alfv\'enic wave.

The fields determined by equations (\ref{eq:totBfield}) and  (\ref{eq:efield}) are used in equation (\ref{eq:motion}) to determine the dynamics of each
pickup $O^+$ ion in the ionosheath of Mars.

\subsection{Initial Conditions and Integrator}\label{sec:ics}

For a particular mode in the plane-wave expansion (\ref{eq:planewave}) one expects that if the time-scale ($\sim 1/\omega$) for the field fluctuation to be  about  the same as the time-scale of gyrotropic motion of the ions ($\sim 1/\Omega$, with $\Omega=qB/mc$), the dynamics of the particles will drastically  be affected. On other hand, in order for an Alfv\'enic-type turbulence to
affect the motion of pickup ions it has to modify such motion in the
time scale spent for the ion to transverse our region of interest; e.g.,
from the subsolar point to the terminator. These considerations guide us, as well as {\sl in situ} measurements, to adopt some numerical values for different parameters in our calculations.

As our fiducial model for the ionosheath of Mars we adopt the following set of values for different parameters in our model.  The range in magnetic field is taken as  $(B_{min},B_{max})=(4, 7)B_0$, where $B_0=5\,$nT, that we choose based on the data from Bertucci et al.~(2003). The SW plasma velocity is taken as $V_{SW}=400\,$km/s and the Alfv\'en velocity of waves $v_A=0.1 \,V_{SW}$. The range of velocities of the SW plasma flow around the planet is assumed to be  $(U_{min},U_{max}) = (0.01,0.5) V_{SW}$, consistent with values measured in that region; see for example the data in Figure~\ref{fig:aspera}.   
Under these conditions,  the mean local cyclotron frequency for a proton in our region of interest is $f_p = \Omega_p/2\pi = 0.42\,$Hz; where the mean field is taken to be  $\langle B \rangle = 5.5 B_0$.  The range of frequencies for the turbulence spectrum is taken from $f_{min} = f_p/16\approx 0.03\,$Hz to $f_{max}=30 f_p\approx 13\,$Hz, that is consistent with the  range in $P(k)$ measured in the ionosheath of Mars (e.g., see Figure~5b in Grebowsky et~al.~2004). For the $v_A$ chosen, the following range in length-scales ($\lambda=2\pi/k$) follows $(\lambda_{min},\lambda_{max})\approx (3.2, 1526)\,$km.

The turbulence spectrum index is taken to be $\gamma\!=\! 2$ in concordance with measurements in the ionosheath of Mars. The amplitude of the turbulent component of the magnetic field is parameterized by the energy ratio $\varepsilon \!=\! \langle \delta B^2 \rangle/B_0^2 $, and we have considered the particular values $\varepsilon = (0.0,0.1,0.5,1.0)$.

The equation of motion (\ref{eq:motion}) for each ion was solved using a  fourth order Runge-Kutta algorithm with an adaptive time-step instead of the standard  Buneman-Boris algorithm (e.g.~Birdsall \& Langdon~1985), a time-symmetric  second-order scheme, used in plasma dynamics. This was adopted after several numerical tests with both integrators, and by the requirement of following more accurately the trajectories of particles  especially when a large amplitude of turbulence ($\varepsilon \approx 1$) and rapid oscillations were  present in the Martian ionosheath. 

 A total of $10^5$ charged particles were used for each value of $\varepsilon$ and propagation direction of turbulent waves considered, yielding a total of one million particle integrations in our numerical experiments. The $10^5$ particles were obtained from sets of $10^4$ particles evolved in 10 different random realizations of the power spectrum $P(k)$, using each time a different initial seed for the random number generator required to construct the turbulent $\delta \mathbf{B}$.  Particles were distributed spatially only in the ``upper'' day-side of Mars, and following an exponential density decay with altitude as in RAP10.

\section{Results}

In this section we present results obtained for the motion of $O^+$ ions under different conditions of turbulence.  We first present the space trajectories of the ions and afterwards their velocity profiles at the terminator.

\subsection{Ion Trajectories}

\begin{figure}[!t]
\centering
\includegraphics[width=0.49\columnwidth]{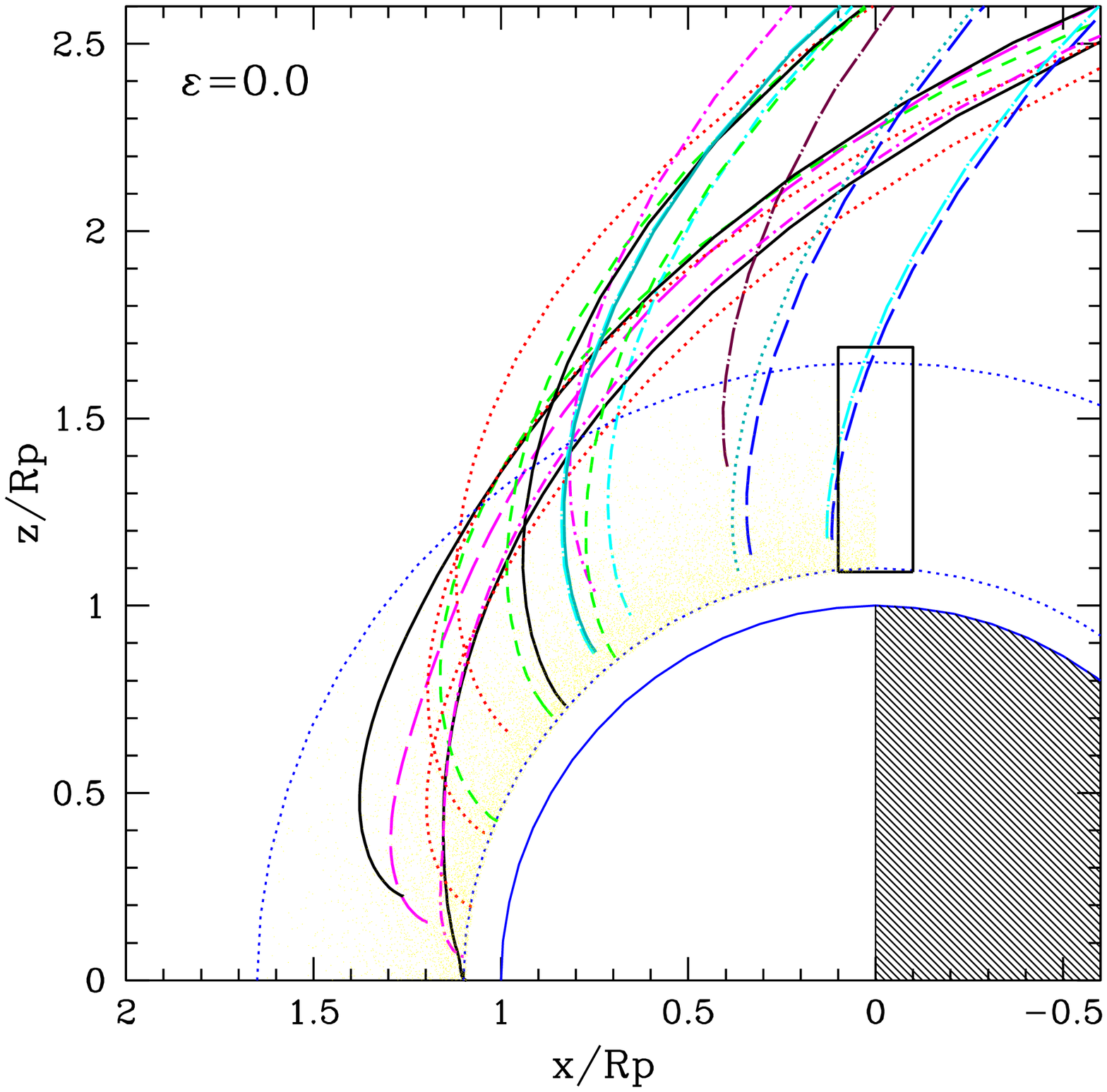}
\includegraphics[width=0.49\columnwidth]{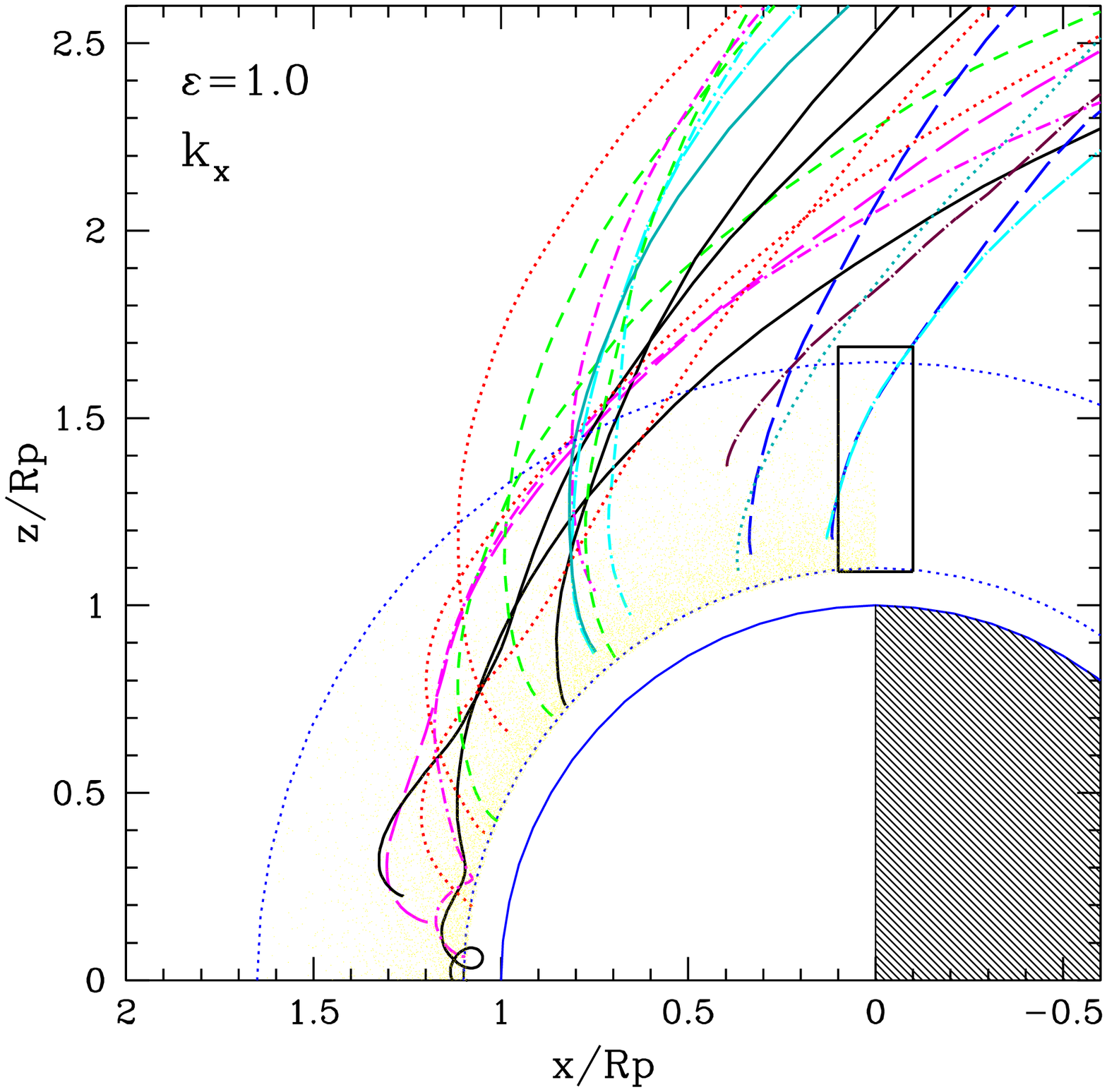}
\includegraphics[width=0.49\columnwidth]{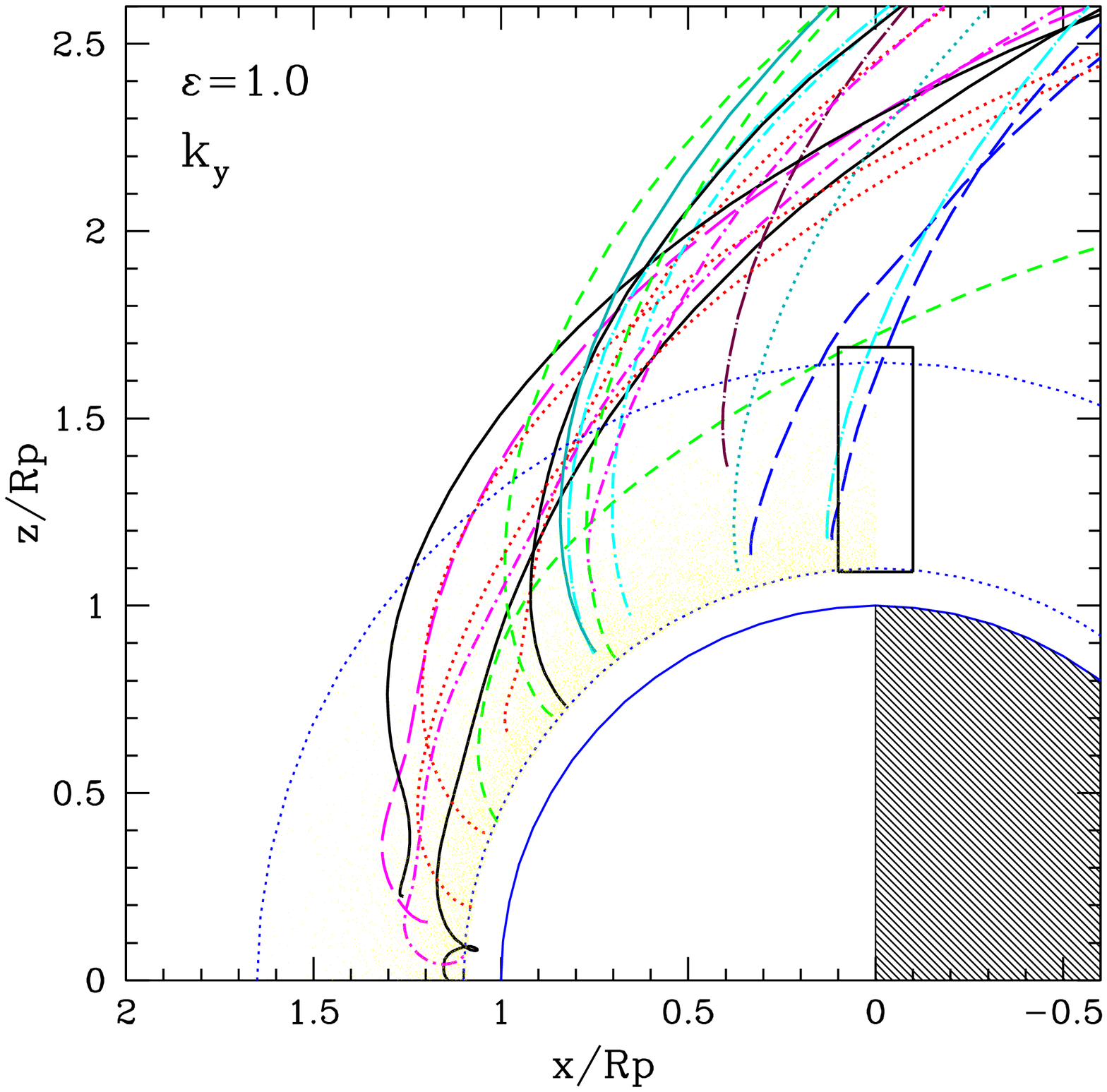}
\includegraphics[width=0.49\columnwidth]{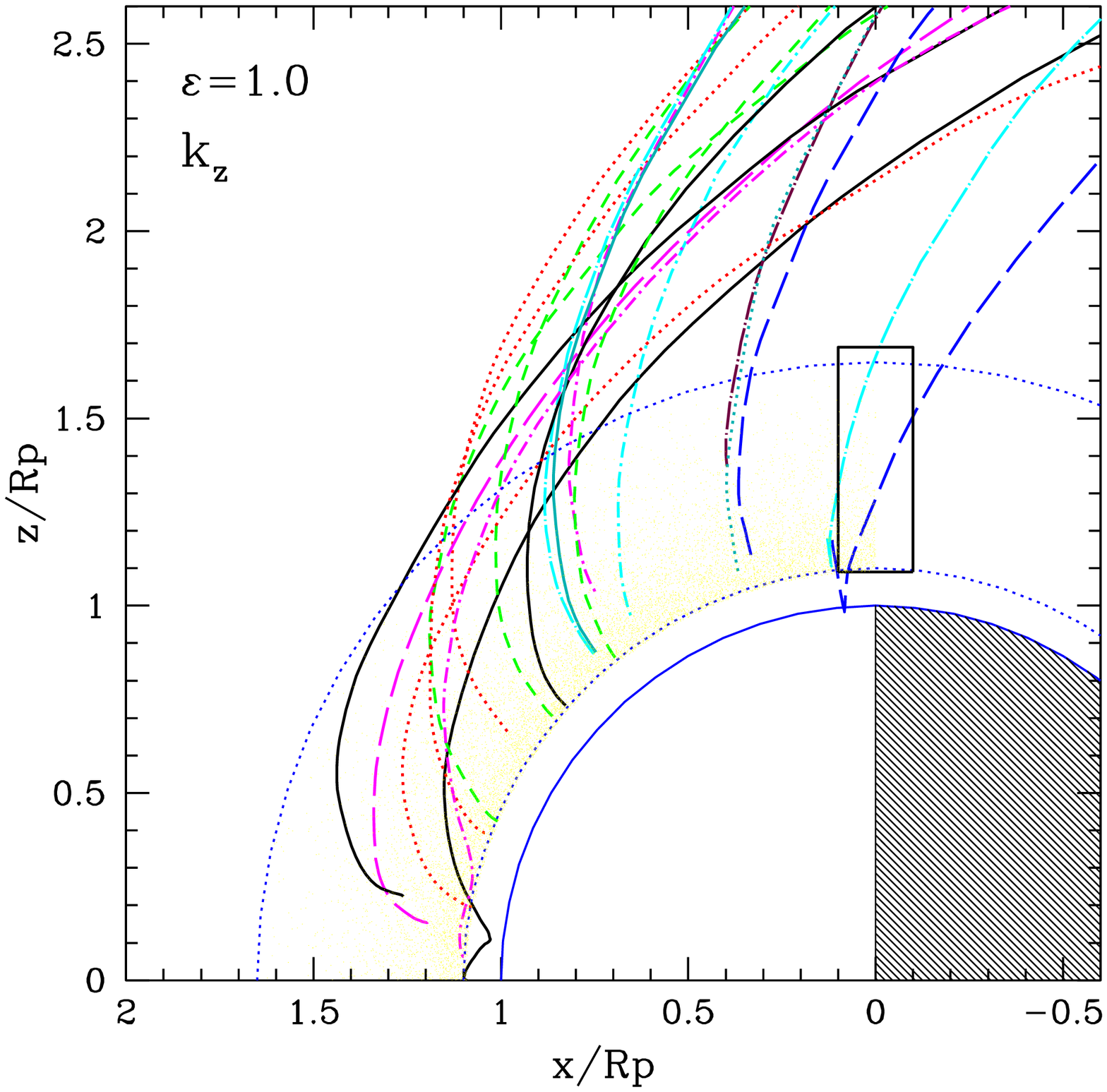}
\caption{Trajectories of $O^+$ ions in a model of the ionosheath of Mars in the $XZ$--plane. Only a turbulence power of $\varepsilon=0$ and $1$ has been considered here. The direction of propagation of turbulence of the wave-vector $k$ is indicated at each panel. }
\label{fig:pathsXZ}
\end{figure}

\begin{figure}[!t]
\centering
\includegraphics[width=0.49\columnwidth]{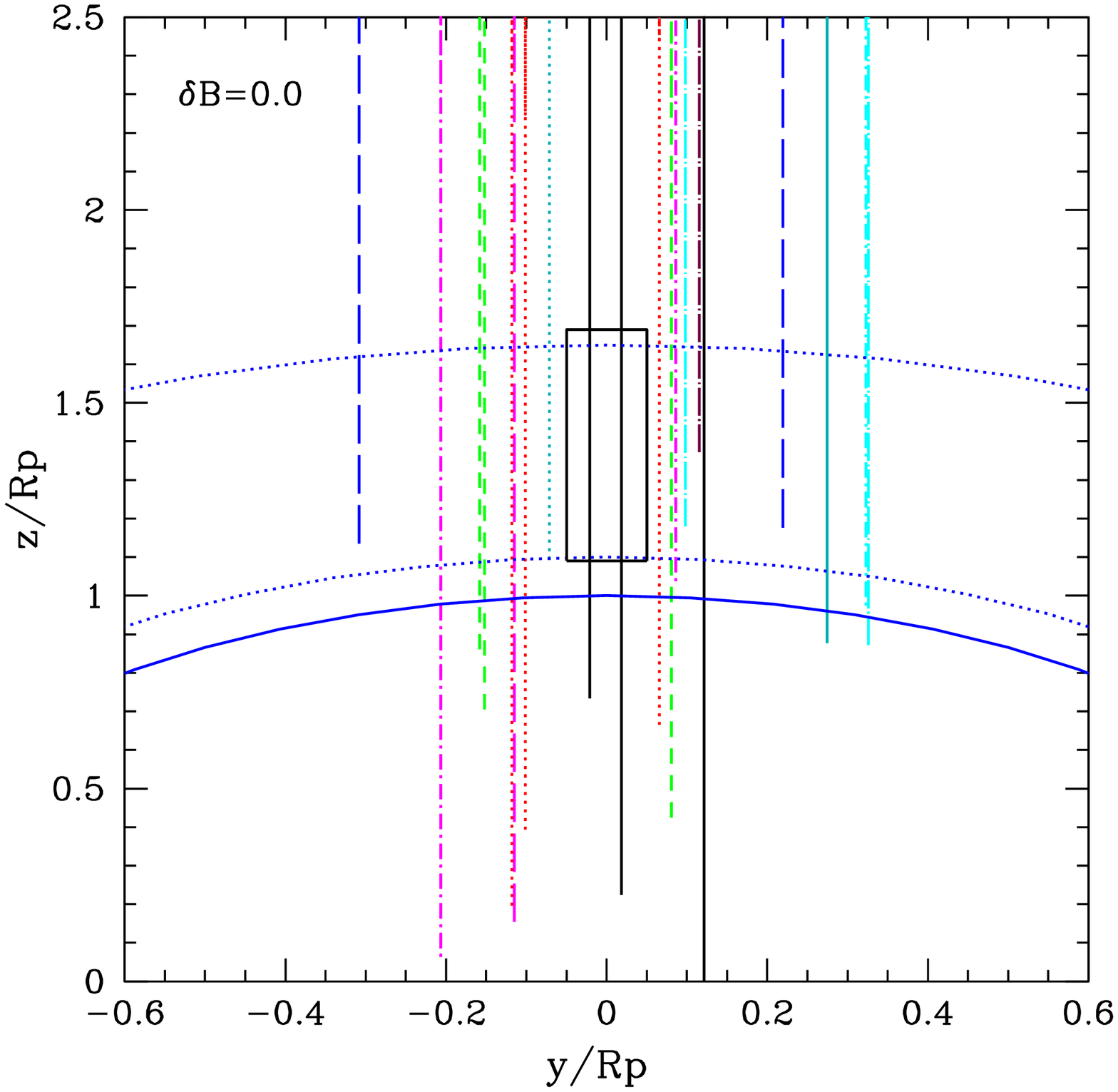}
\includegraphics[width=0.49\columnwidth]{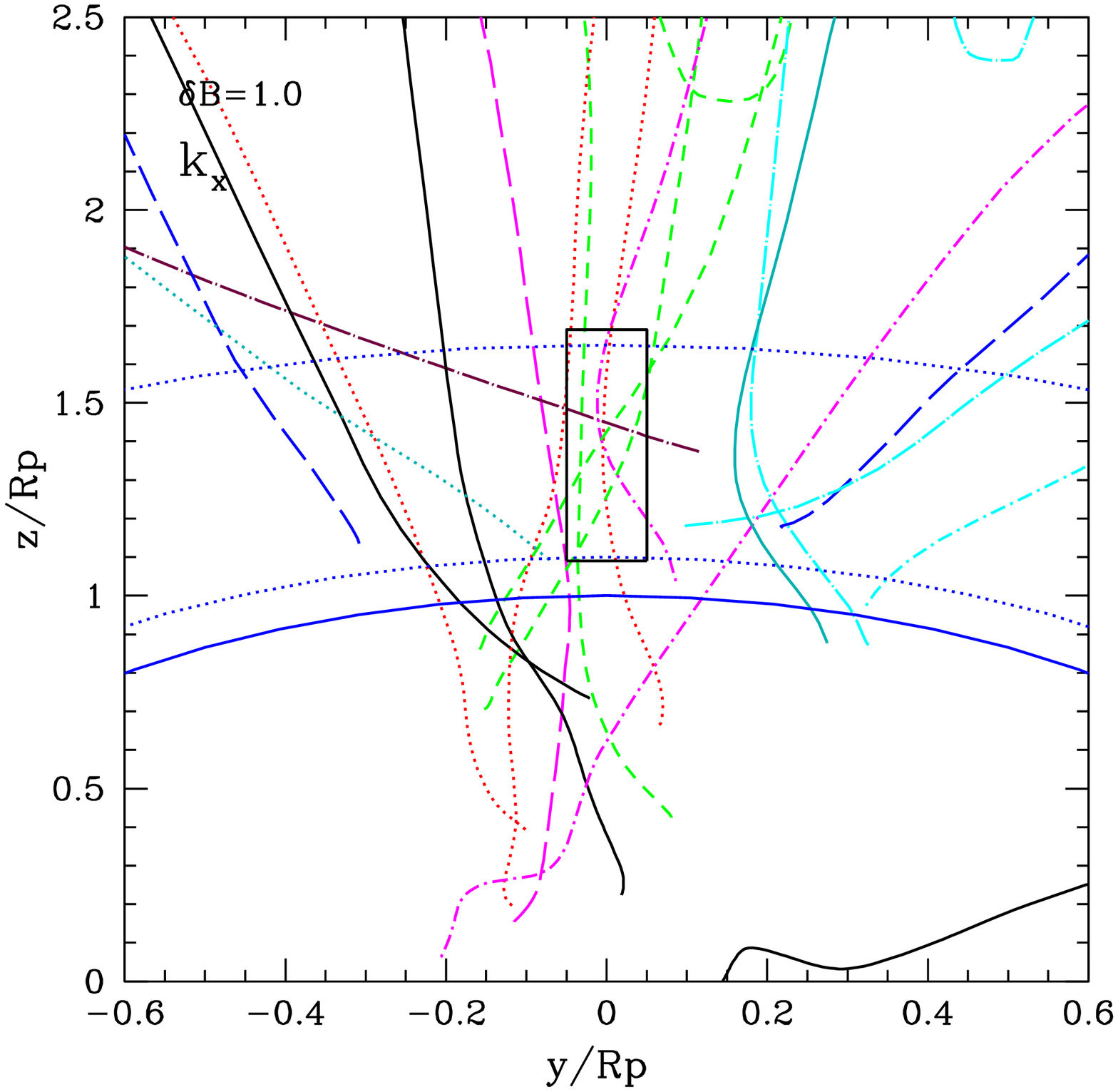}
\includegraphics[width=0.49\columnwidth]{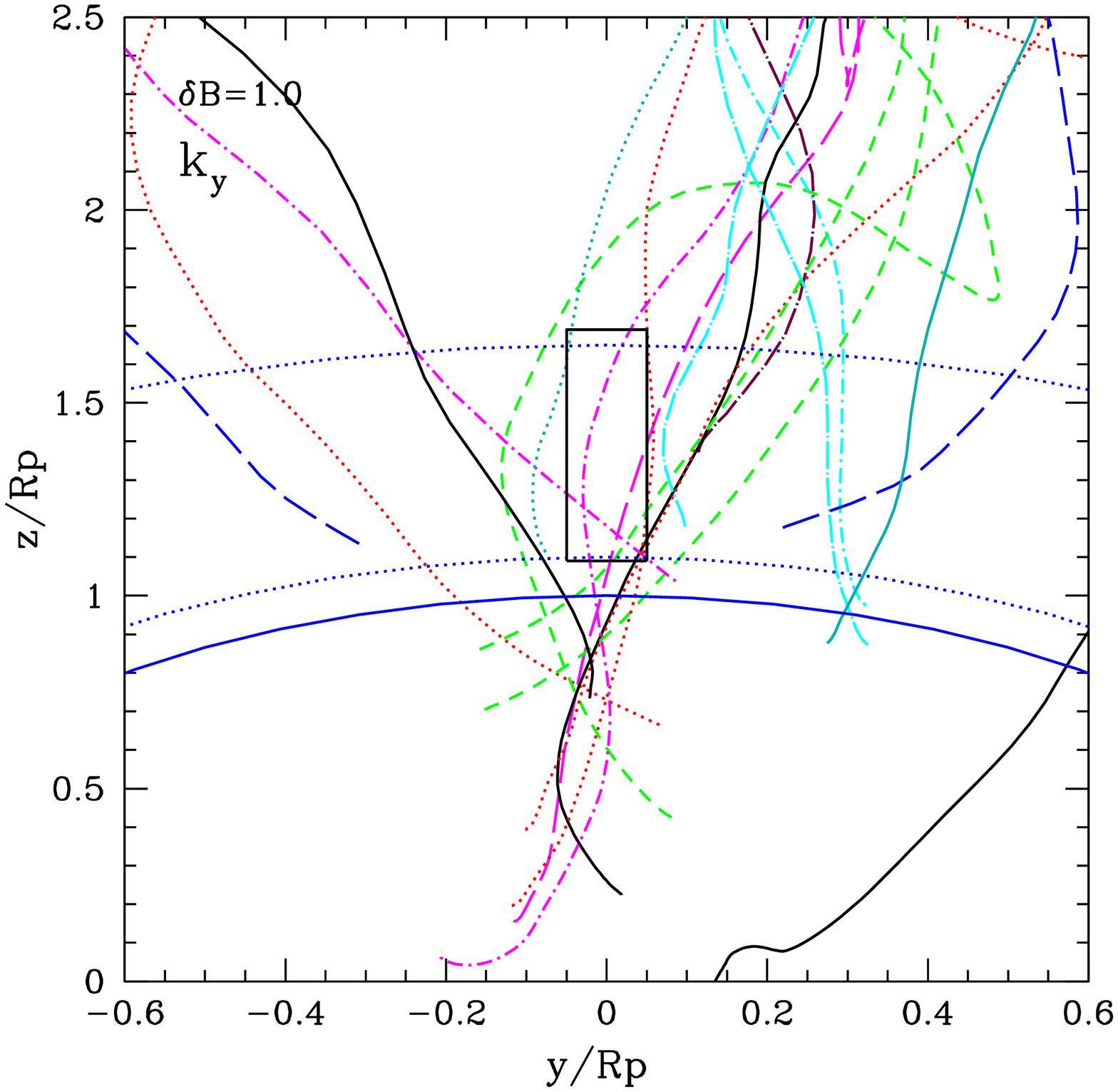}
\includegraphics[width=0.49\columnwidth]{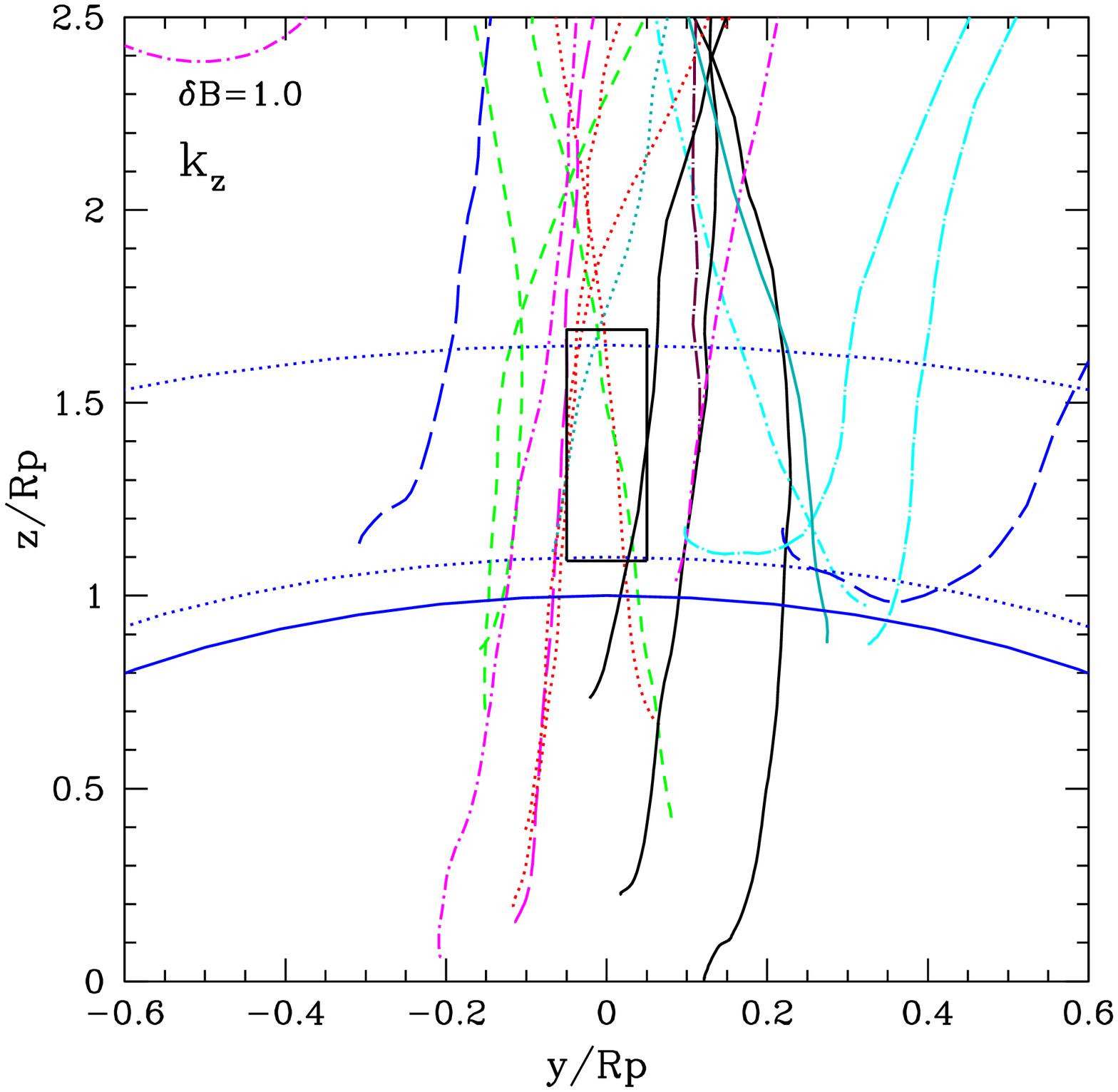}
\caption{As in Figure~\ref{fig:pathsXZ} but for motion projected on the $YZ$--plane.}
\label{fig:pathsYZ}
\end{figure}

In Figure~\ref{fig:pathsXZ} we show twenty random trajectories of $O^+$ ions in the $XZ$ plane for turbulent waves propagating in the $x$, $y$ and $z$-direction for the extreme amplitude $\varepsilon \! =\! 1$, along with the $\varepsilon \!=\! 0$ case, and in  Figure~\ref{fig:pathsYZ} the corresponding ones in the $YZ$-plane. The square box in these figures correspond approximately to an altitude range from $300\,$km to $2300\,$km; see Figure~\ref{fig:aspera}.

The motion of pickup oxygen ions without turbulence ($\varepsilon\!=\!0.0$), is qualitatively similar to that found in previous works on the dynamics of pickup ions with an analytical model for the $\mathbf{B}$-field (e.g. Luhmann \& Schwingenschuh~1990, Luhmann~1990, Kallio \& Koskinen~1999, RAP10) and from that derived from MHD simulations (e.g. Jin et~al.~2001, B{\"o}{\ss}wetter et~al.~2007, Fang~et~al.~2008, Fang~et~al.~2010). 

The presence of even a small amplitude ($\varepsilon=0.1$) of turbulent waves propagating in the ionosheath shows a strong effect in the $YZ$-motion of pickup ions, while it is less noticeable in the $XZ$-plane. Even in the case of $\varepsilon=1.0$ the  $XZ$-motion is not greatly affected (Figure~\ref{fig:pathsXZ}), in particular the trajectories are not importantly  ``bent'' toward the SW  direction of motion as is inferred from the Mars-Express measurements (Figure~\ref{fig:aspera}).

\begin{figure*}[!t]
\centering
\includegraphics[width=0.31\textwidth]{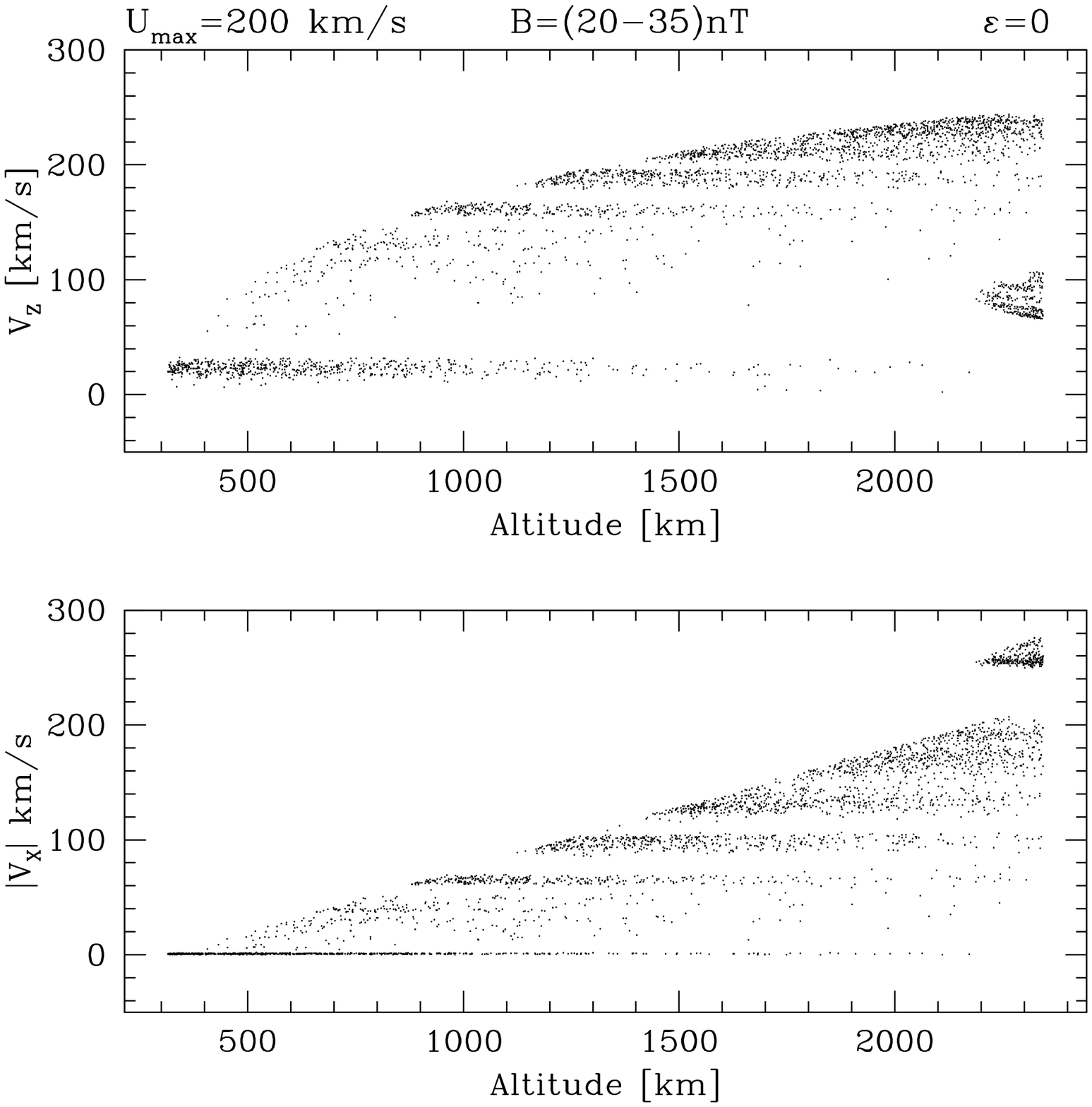}
\hspace{2pt}
\includegraphics[width=0.31\textwidth]{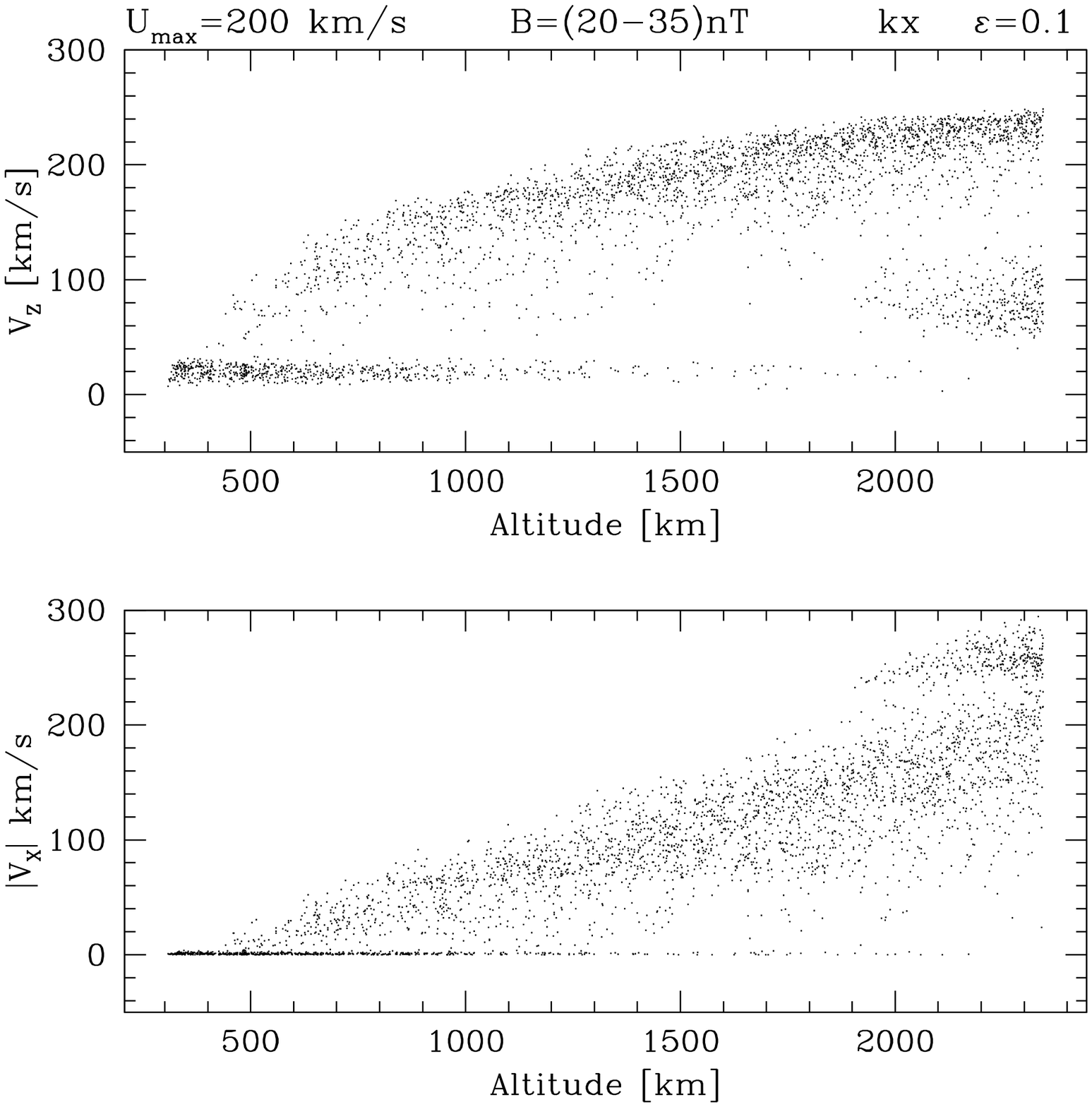}
\hspace{2pt}
\includegraphics[width=0.31\textwidth]{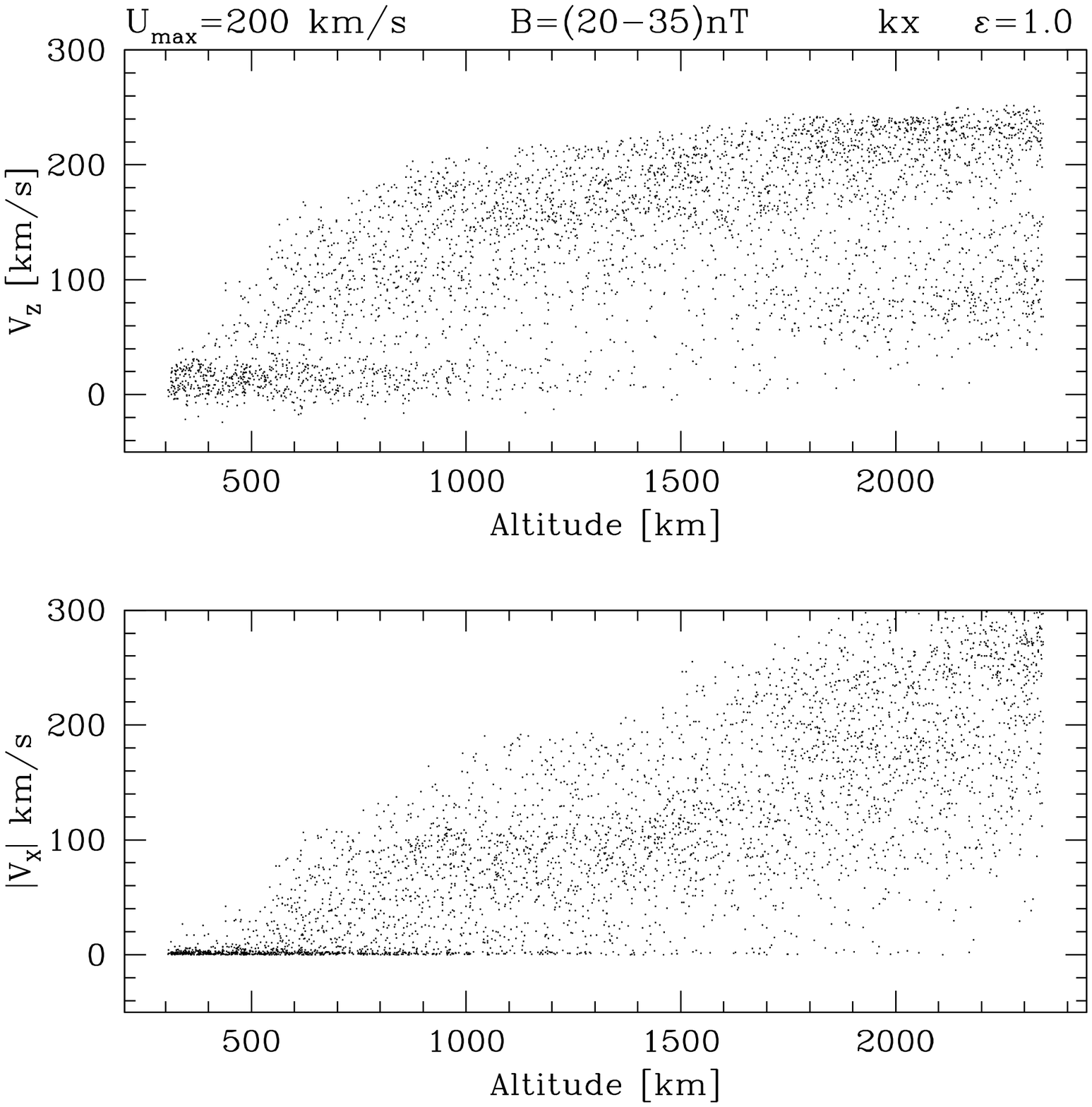}
\caption{Phase-space diagrams of $O^+$ ions at the Martian terminator: $V_z-z$ and  $|V_x|-z$. Turbulent waves propagating in the $X$-direction of increasing amplitudes $\varepsilon$ are shown from left to right.}
\label{fig:phase}
\end{figure*}

 The dynamics of pickup ions is clearly very complex even under a small ``perturbation'' to the background field that, for example, may preclude any estimation of the ionosheath conditions based solely in the analysis of the motion of pickup ions. The analysis of the statistical motion of the ions is out of the scope of this paper, but we expected that they will have a dynamics similar to that of a L\'evy random walk process  (e.g. Shlesinger, West \& Klafter~1987, Greco~et~al.~2003).

\subsection{Velocity Profiles}

Several processes may be investigated from the velocities of the ions in the turbulent medium, such as their phase-space structure or the pitch-angle difussion and scattering (e.g.~Price \& Wu~1987,  Li~et~al.~1997, Cravens~et~al.~2002, Fang~et~al.~2008). Although important for our understanding of $O^+$ pickup ions, our focus here is essentially in the vertical velocity profiles of the ions over the terminator in order to compare with the data shown in Figure~\ref{fig:aspera}.

The vertical velocity profiles of $O^+$ ions were computed at the terminator,  and considering only a small region of width in the $Y$--direction of $\Delta y=\pm R_M/4$; as indicated in Figure~\ref{fig:pathsYZ}. Twenty bins in the $Z$--direction were set and the average velocity of all particles crossing the terminator were computed for each component. This was done for each level of turbulence and direction of the Alfv\'enic waves.

In Figure~\ref{fig:phase} we show ``phase-space'' plots of the particles above the Martian terminator, in the box described above, for our fiducial parameters. The different graphs correspond different levels of turbulence ($\varepsilon=0,0.1,1$) propagating in the $X$--direction. A clear effect of heating and scattering in phase-space due to turbulence is appreciated as its amplitude $\varepsilon$ increases in value. Under our fiducial values, and the different prescriptions for turbulent waves, we notice already from  Figure~\ref{fig:phase} that the $Z$--component of the velocity dominates over the $X$--component.

\begin{figure*}[!t]
\centering
\includegraphics[width=0.31\textwidth]{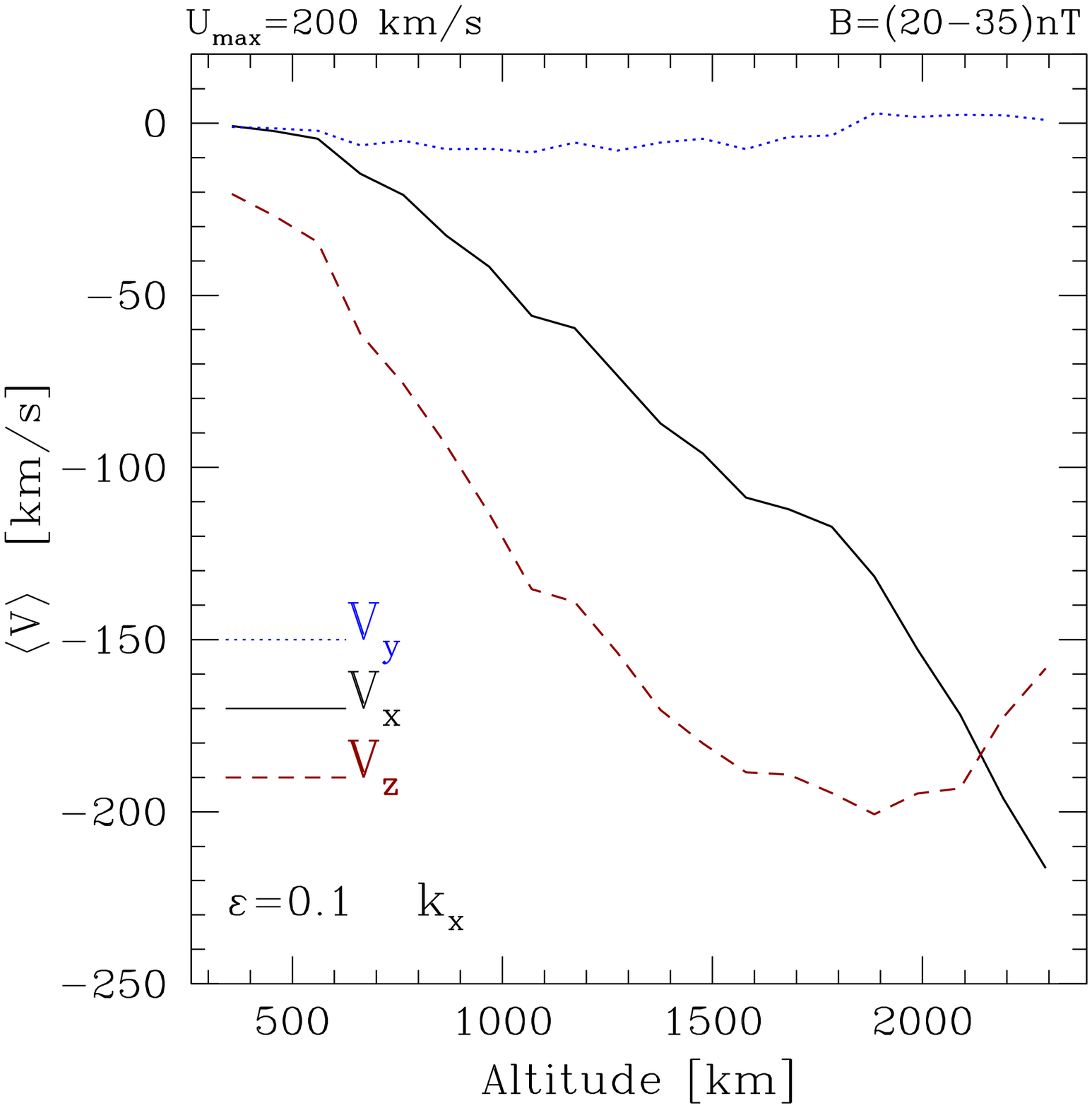}
\hspace{2pt}
\includegraphics[width=0.31\textwidth]{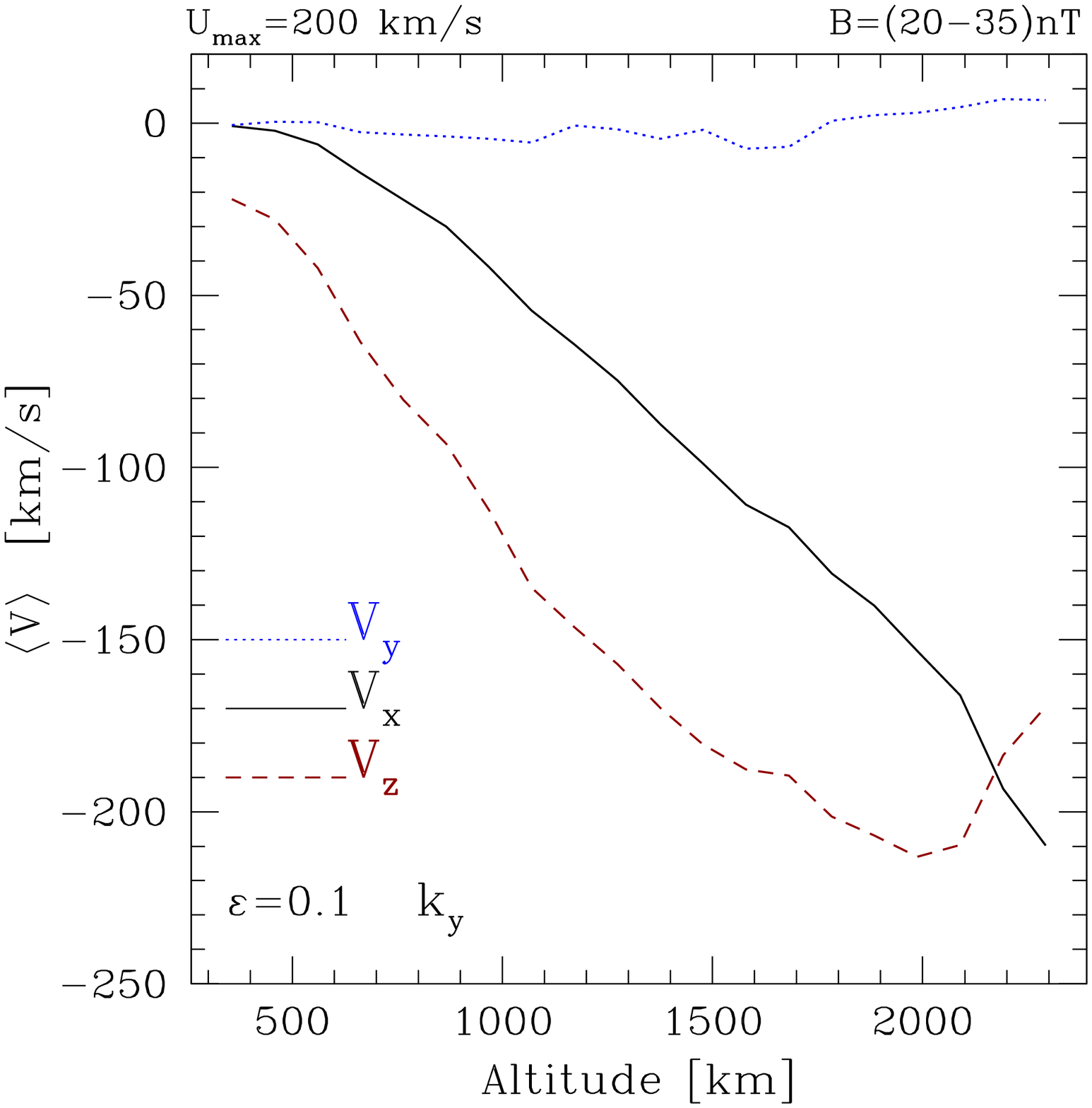}
\hspace{2pt}
\includegraphics[width=0.31\textwidth]{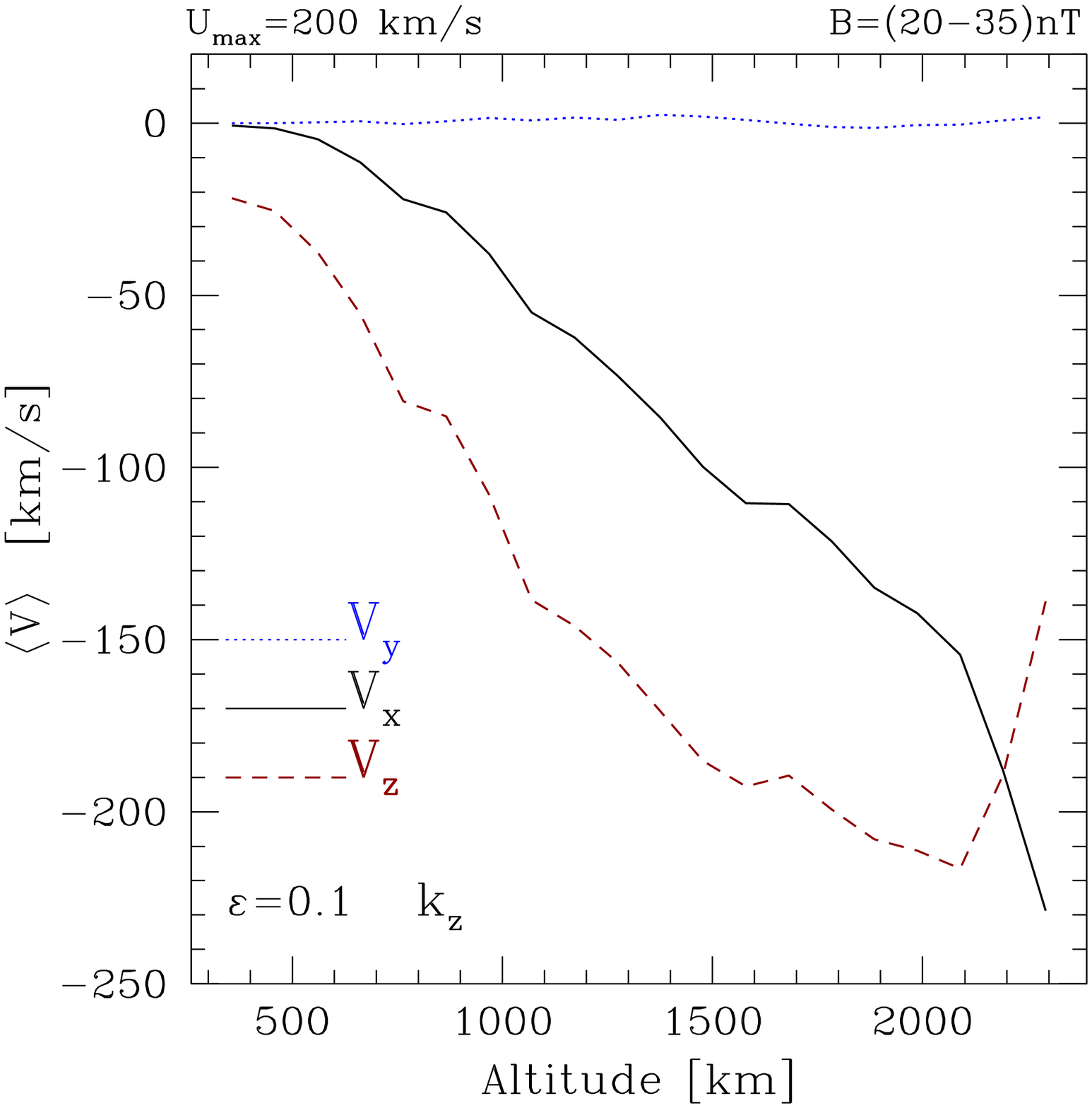}
\hspace{2pt}
\includegraphics[width=0.31\textwidth]{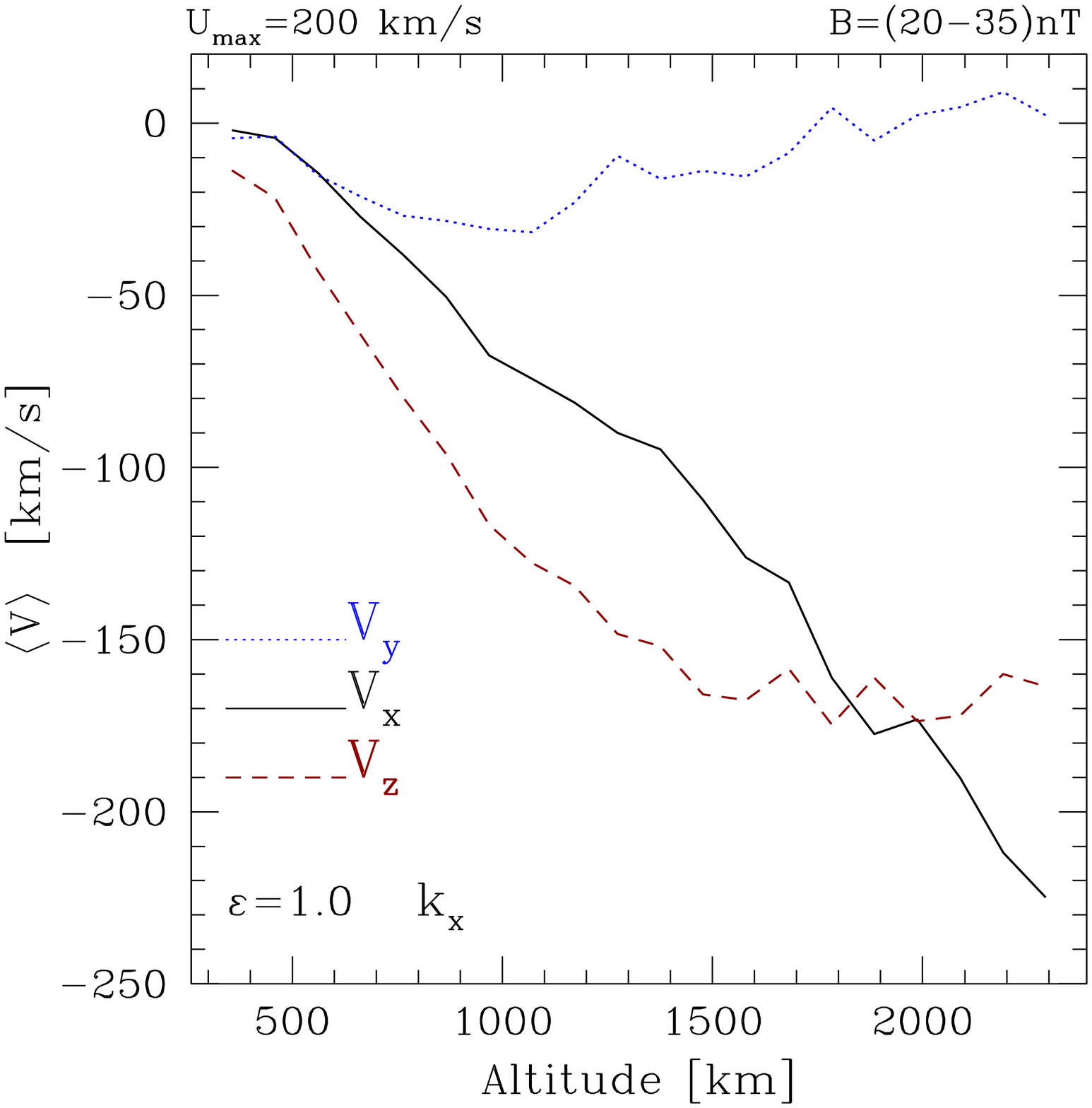}
\hspace{2pt}
\includegraphics[width=0.31\textwidth]{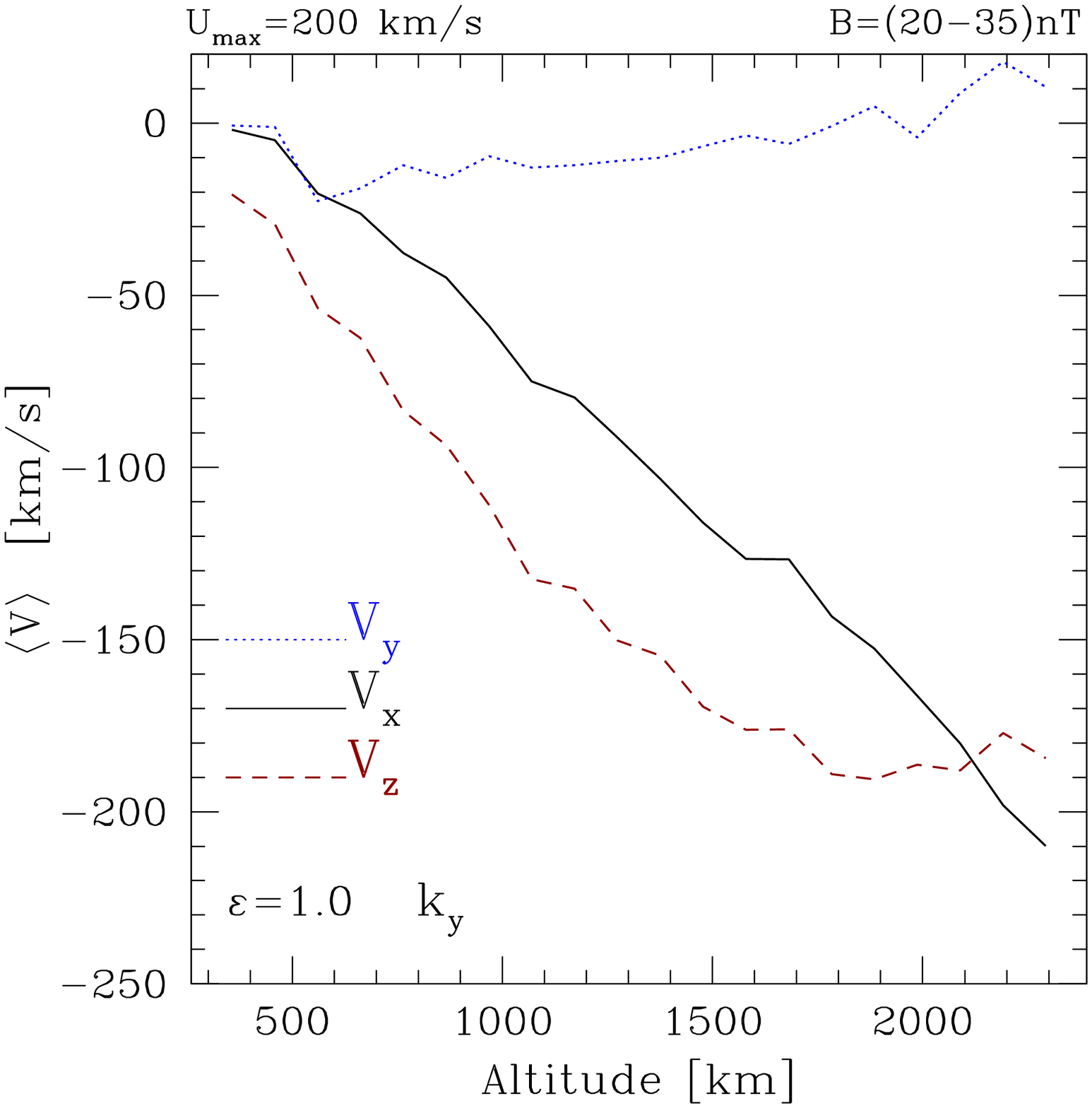}
\hspace{2pt}
\includegraphics[width=0.31\textwidth]{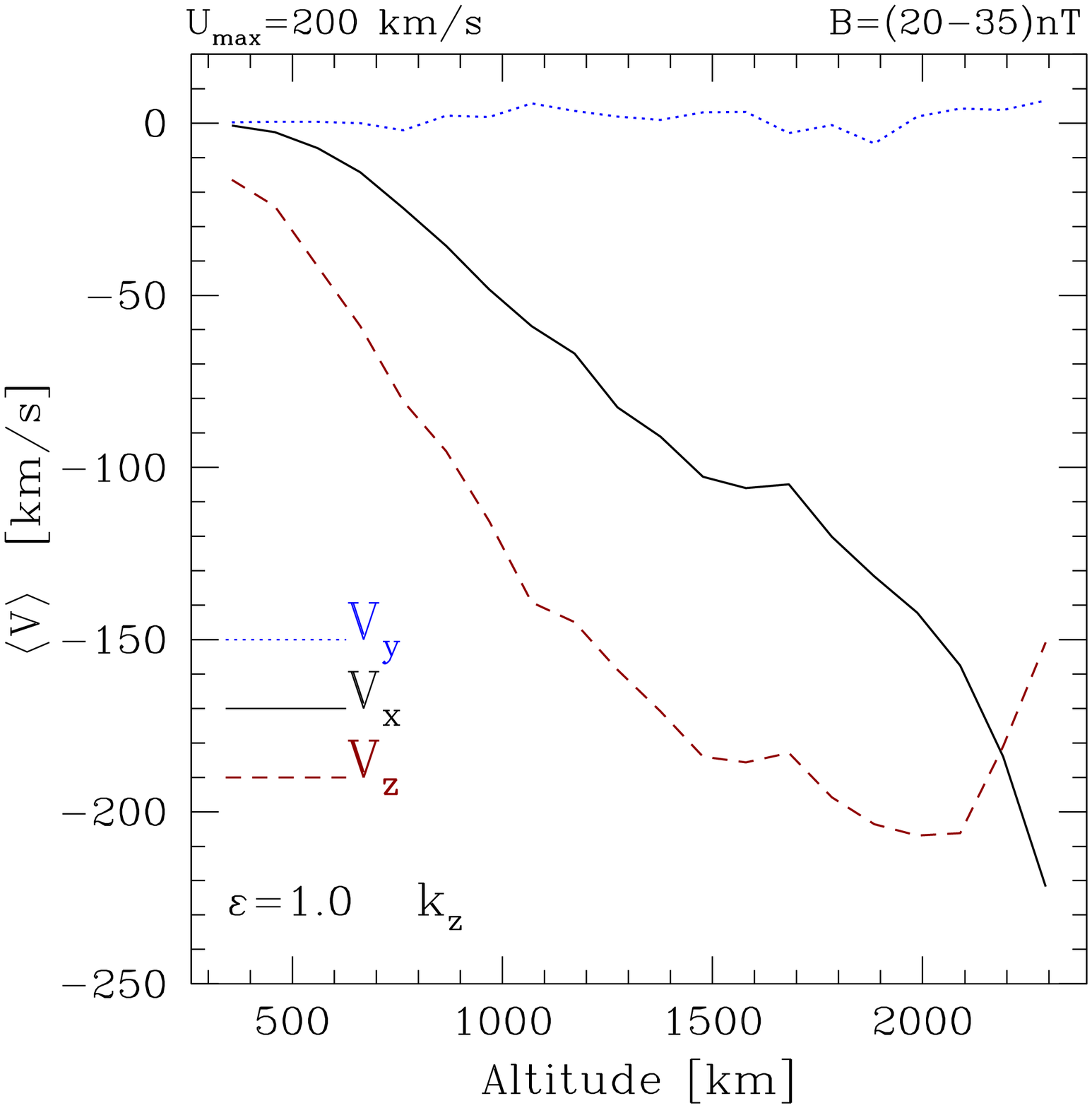}
\caption{Vertical velocity profiles of $O^+$ ions at the Martian  terminator. From left to right, the plots correspond to different turbulent waves propagation vector. The upper row correspond to values of $\varepsilon=0.1$ and  the lower for $\varepsilon=1.0$. }
\label{fig:profiles}
\end{figure*}

In Figure~\ref{fig:profiles} we show the average velocities of $v_x$, $v_y$ and $v_z$ of the  $O^+$ ions, for turbulence amplitudes  $\varepsilon=0.1$ ({\it top}) and $\varepsilon=1.0$ ({\it bottom}). The case of $\varepsilon=0$ has a similar behaviour as that of $\varepsilon=0$ and we do not consider it in Figure~\ref{fig:profiles}. The graphical representation is similar to that in Figure~\ref{fig:aspera} in order to aid in the comparison with the {\it in situ} data; the $Z$--component of the velocity has been inverted in sign for that purpose.

The trend displayed of  the velocity profiles is similar to that reported in RAP10, although significant differences appear now when turbulent waves are taken into account. In the no-turbulent case, $\varepsilon=0$, the $\langle v_y \rangle$ is  zero since no motion along the $Y$--direction is generated (see Figure~\ref{fig:pathsYZ}). Once a turbulent component sets in, the erratic motion of particles contribute to a non-zero value of $\langle v_y \rangle$.

A clear result from all of our numerical calculations is that even under a strong ($\varepsilon=1$) level of turbulent waves propagating in any direction, the trend of the $Z$-component of the velocity of ions always dominating over the $X$-component is preserved, contrary  to measurements at such altitudes in the Martian terminator.  

Unfortunately, for example, the extensive and complete MHD  simulations of Fang et~al.~(2008) do not address the velocity profiles of pickup ions at the terminator, so we are not able to compare our results with their work. In this respect, it will be interesting to measure from this kind of MHD simulations the vertical profiles of $O^+$ ions and compare them with data.

\section{Discussion and Final Comments}

\subsection{Discussion}

We have calculated the trajectory and average velocity profiles of newly born $O^+$ ions in the dayside  ionosheath and magnetic polar regions of Mars, as they are accelerated by the convective electric field due to the streaming solar wind plasma. In addition to a simplified model of the draped IMF, which determines the large-scale geometry of the magnetic field, we have included for the first time an approximation to the ``turbulent'' component for the magnetic field in that region, and studied its effect on the dynamics of pickup ions. We have compared the velocity profiles resulting from our calculations to those measured in the region with the ASPERA-3 instrument onboard the Mars Express, as reported by PdT09.

\begin{figure*}[!t]
\centering
\includegraphics[width=0.31\textwidth]{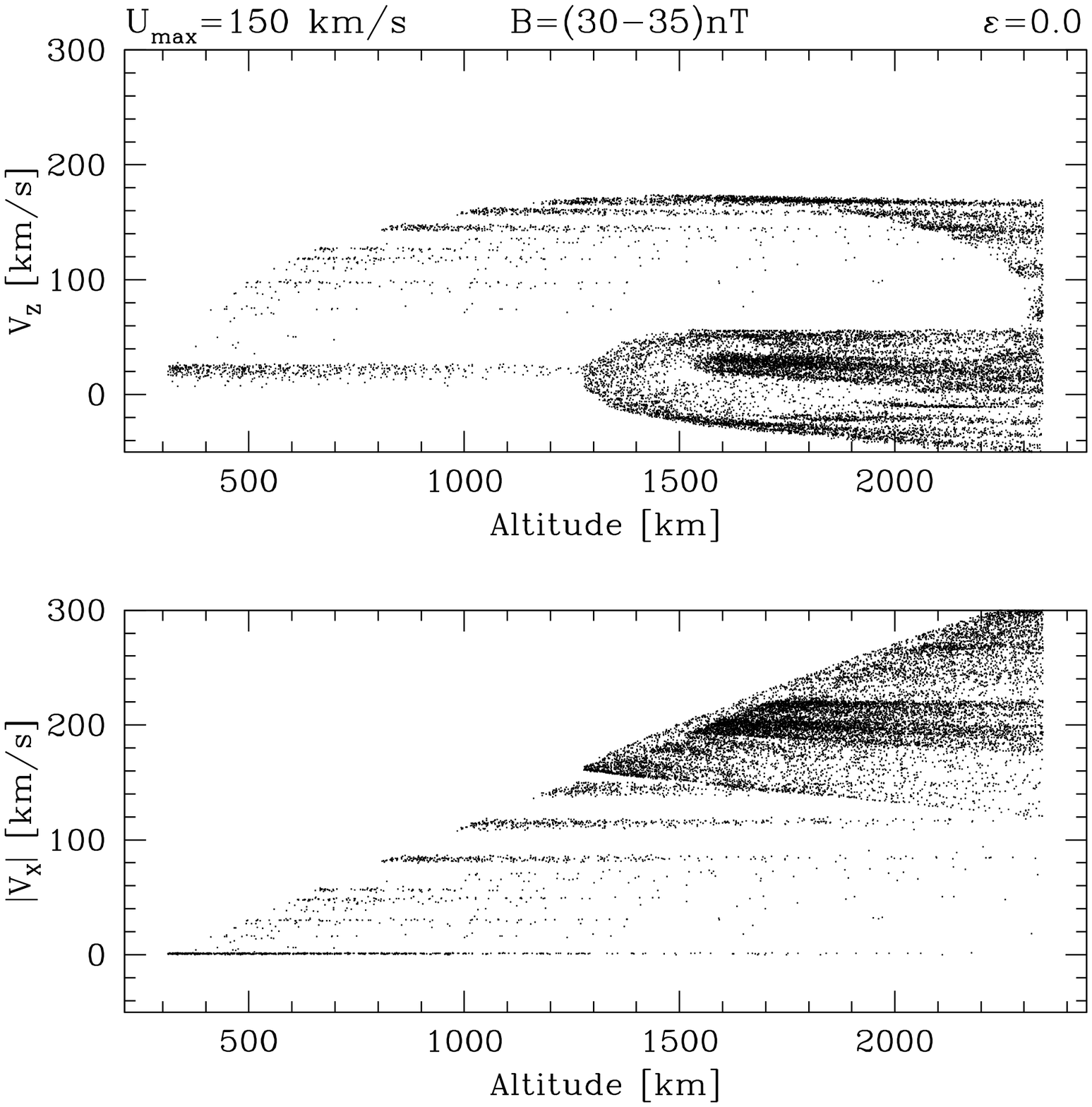}
\hspace{2pt}
\includegraphics[width=0.31\textwidth]{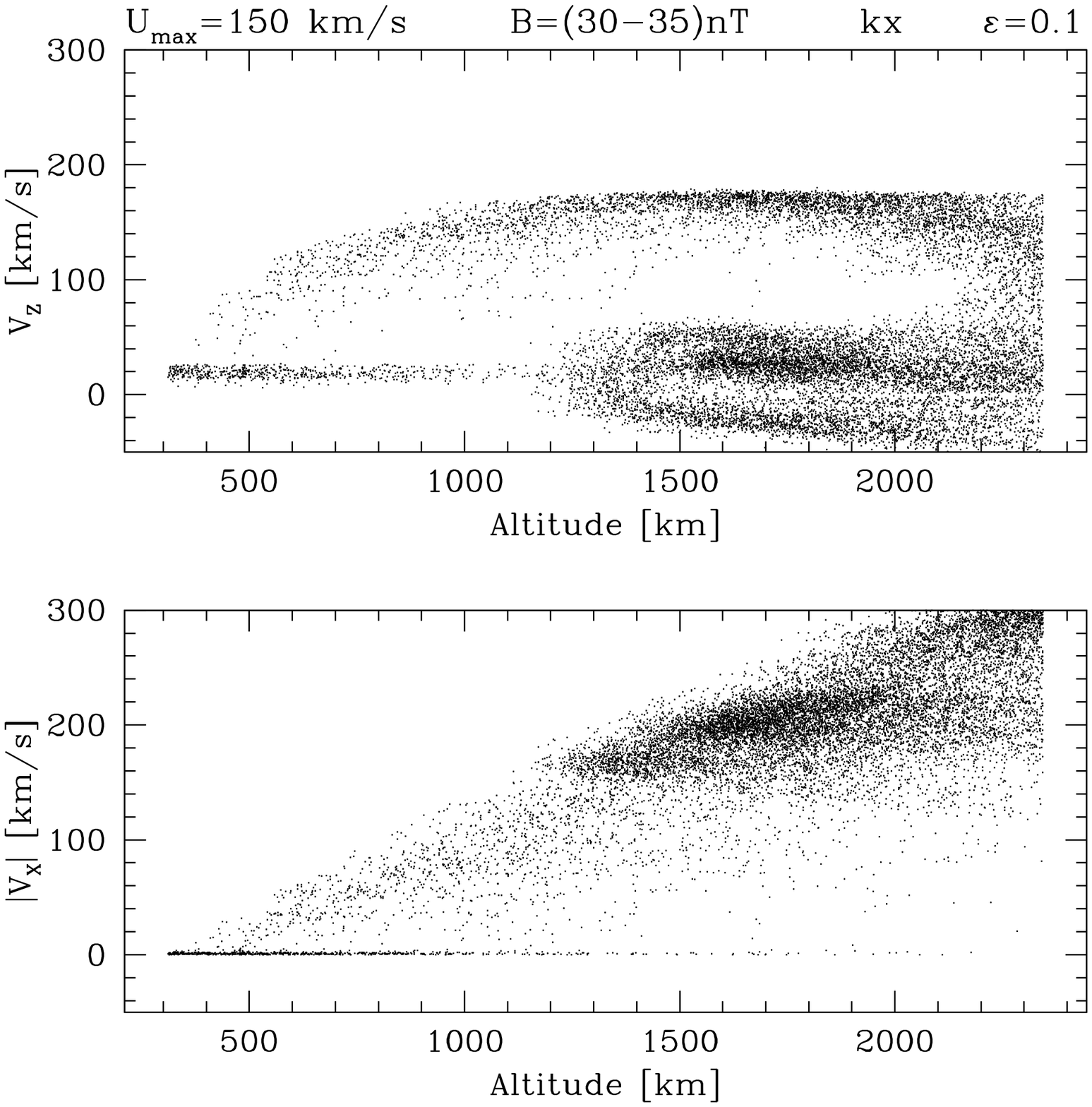}
\hspace{2pt}
\includegraphics[width=0.31\textwidth]{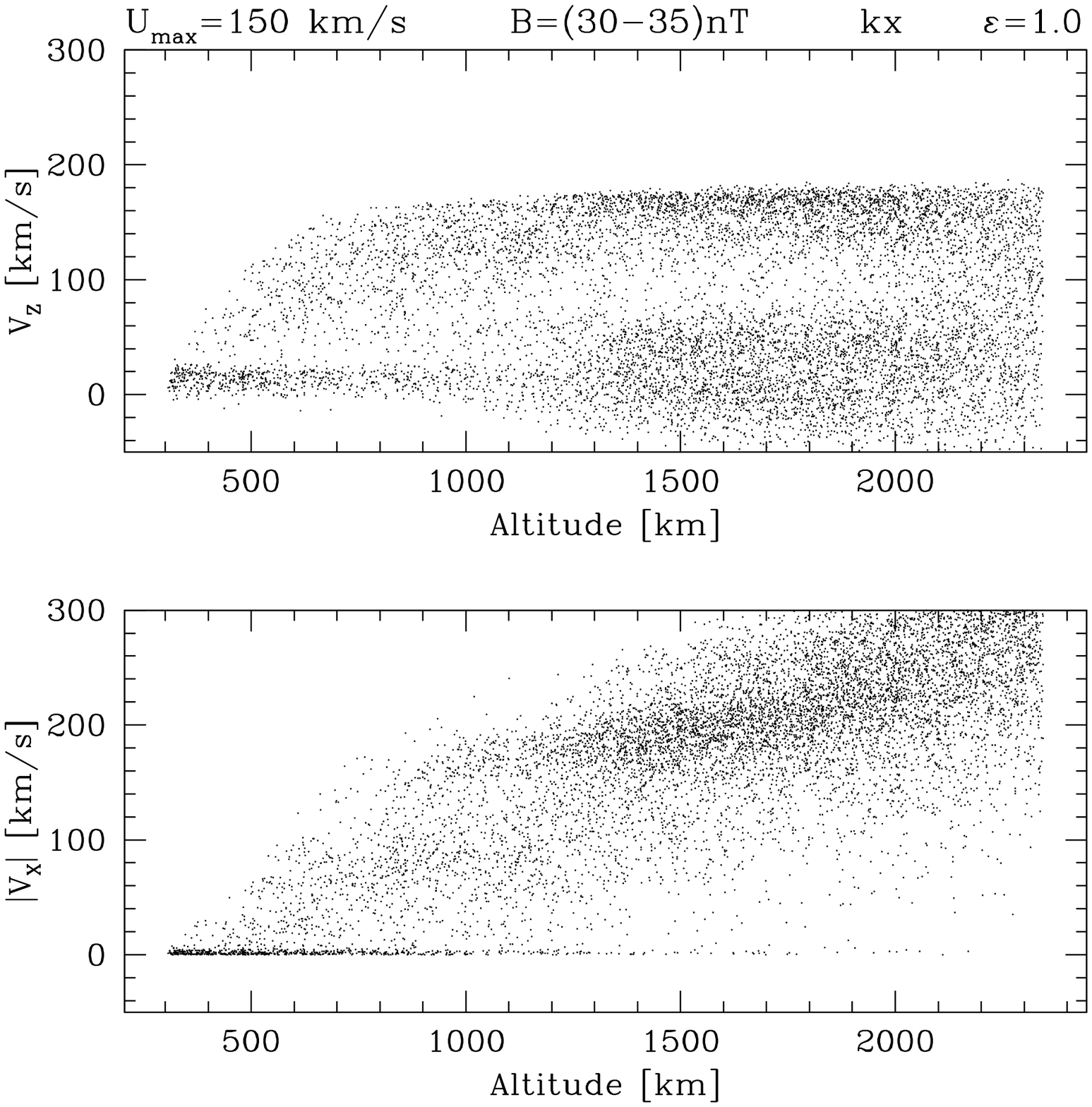}
\caption{The same as in Figure~\ref{fig:phase} but with parameters more  favorable for a strong curvature in the paths of $O^+$ ions. An increase of particle density in different regions of the phase-space is noted in comparison to that ofFigure~\ref{fig:phase}.  }
\label{fig:phaseE}
\end{figure*}

The effect of a turbulent part $\delta \mathbf{B}$ in the magnetic field is most significant in deflecting the trajectories in the plane perpendicular to the SW flow (plane $YZ$), but it has little effect on modifying the velocity structure of pickup ions at the terminator in comparison to the absence of turbulent waves. This behavior holds under different power-laws ($\gamma$) for the turbulence power spectrum and its amplitude ($\varepsilon$), as well as for its direction of propagation and small variations of our fiducial values ($\S$\ref{sec:ics}) of the parameters in the simulations.

\begin{figure*}[!t]
\centering
\includegraphics[width=0.31\textwidth]{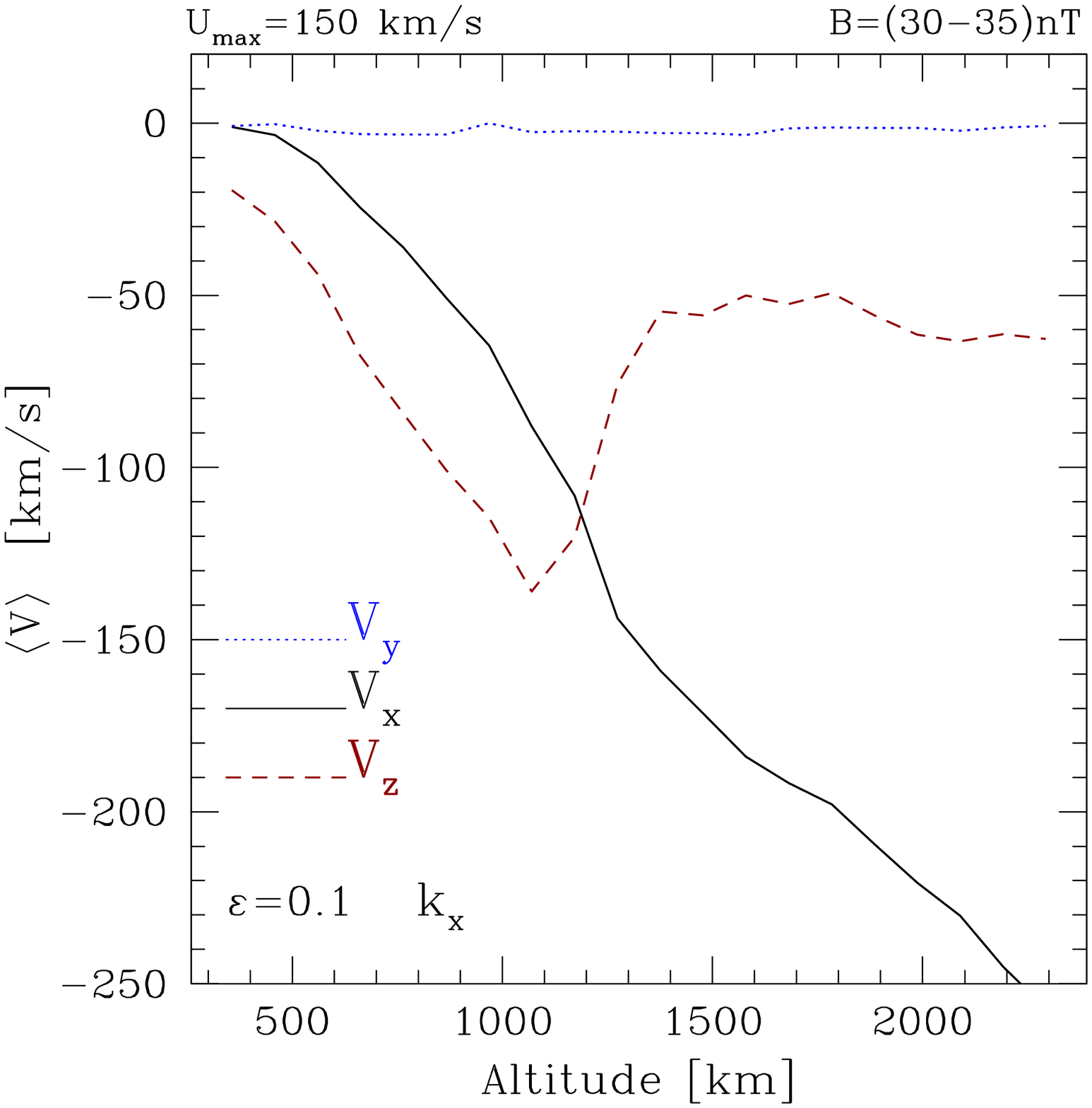}
\hspace{2pt}
\includegraphics[width=0.31\textwidth]{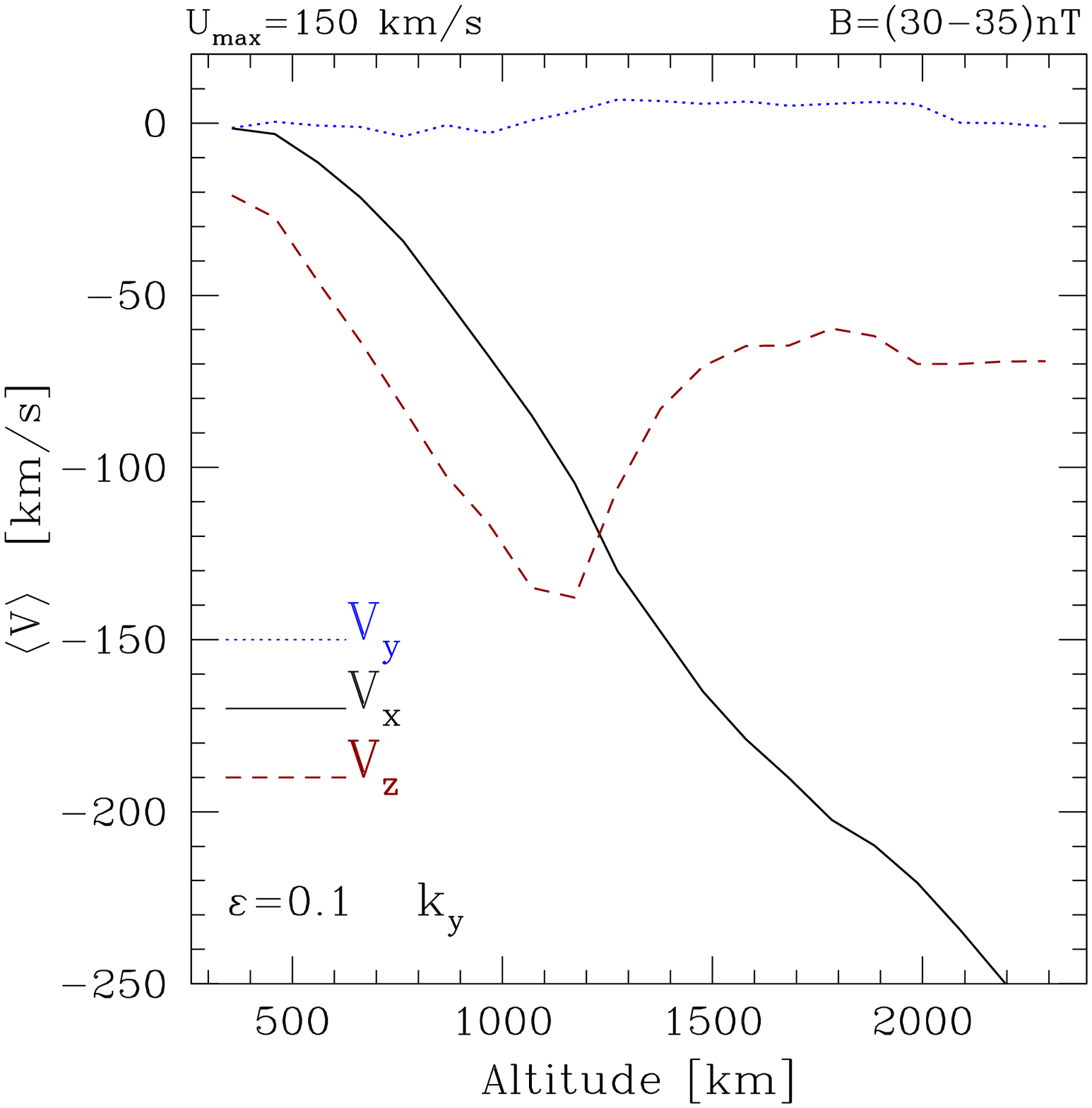}
\hspace{2pt}
\includegraphics[width=0.31\textwidth]{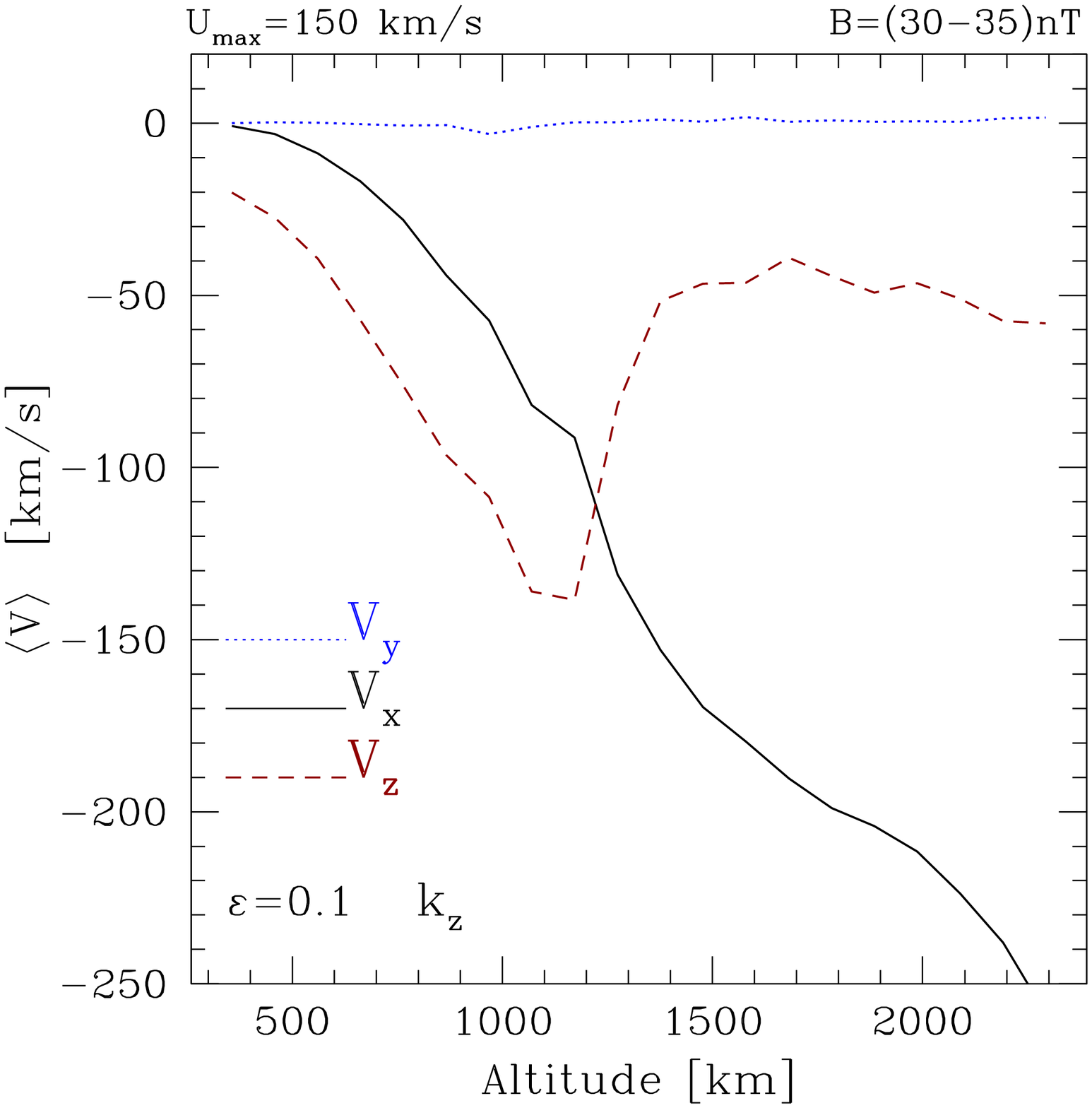}
\hspace{2pt}
\includegraphics[width=0.31\textwidth]{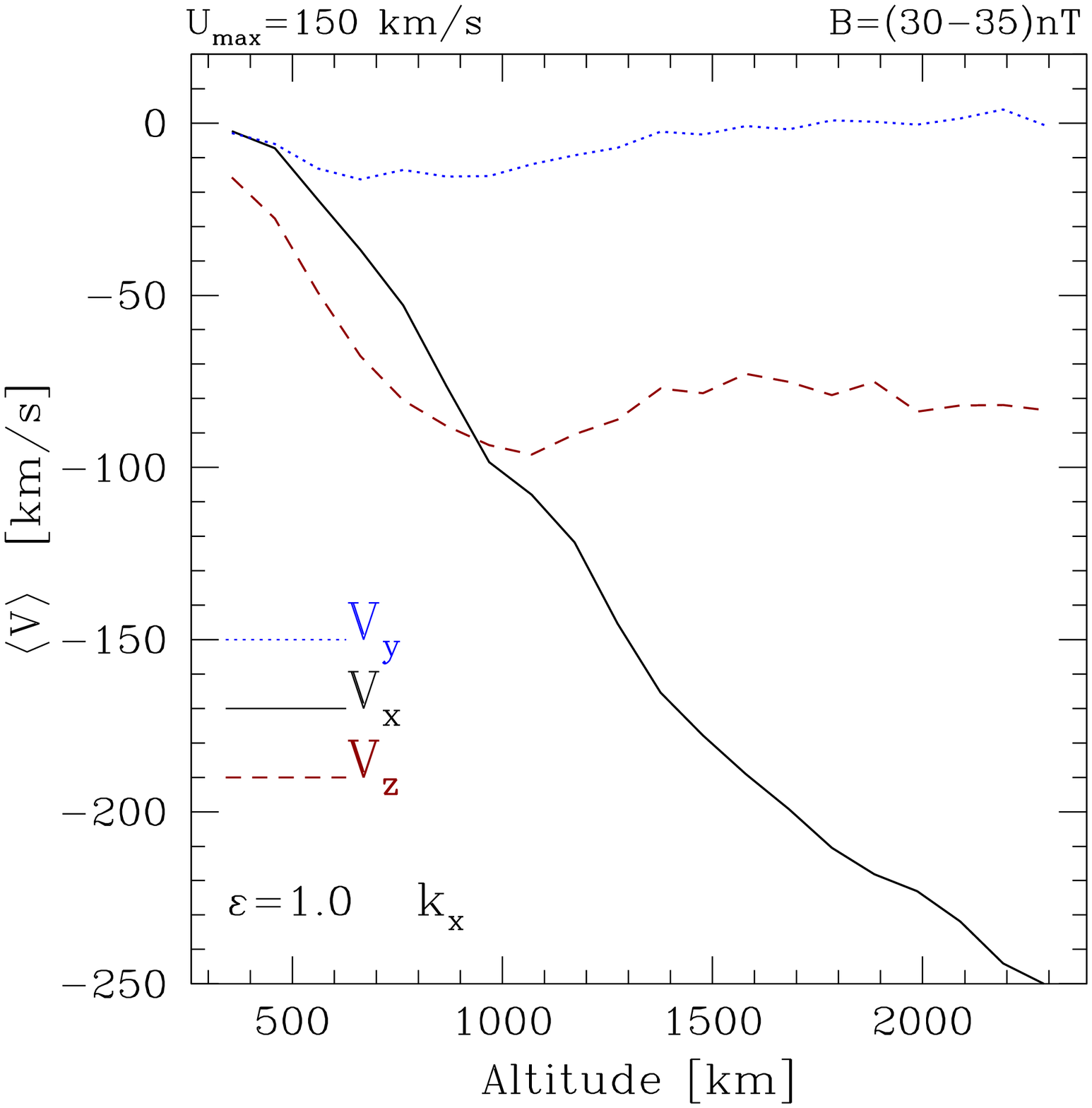}
\hspace{2pt}
\includegraphics[width=0.31\textwidth]{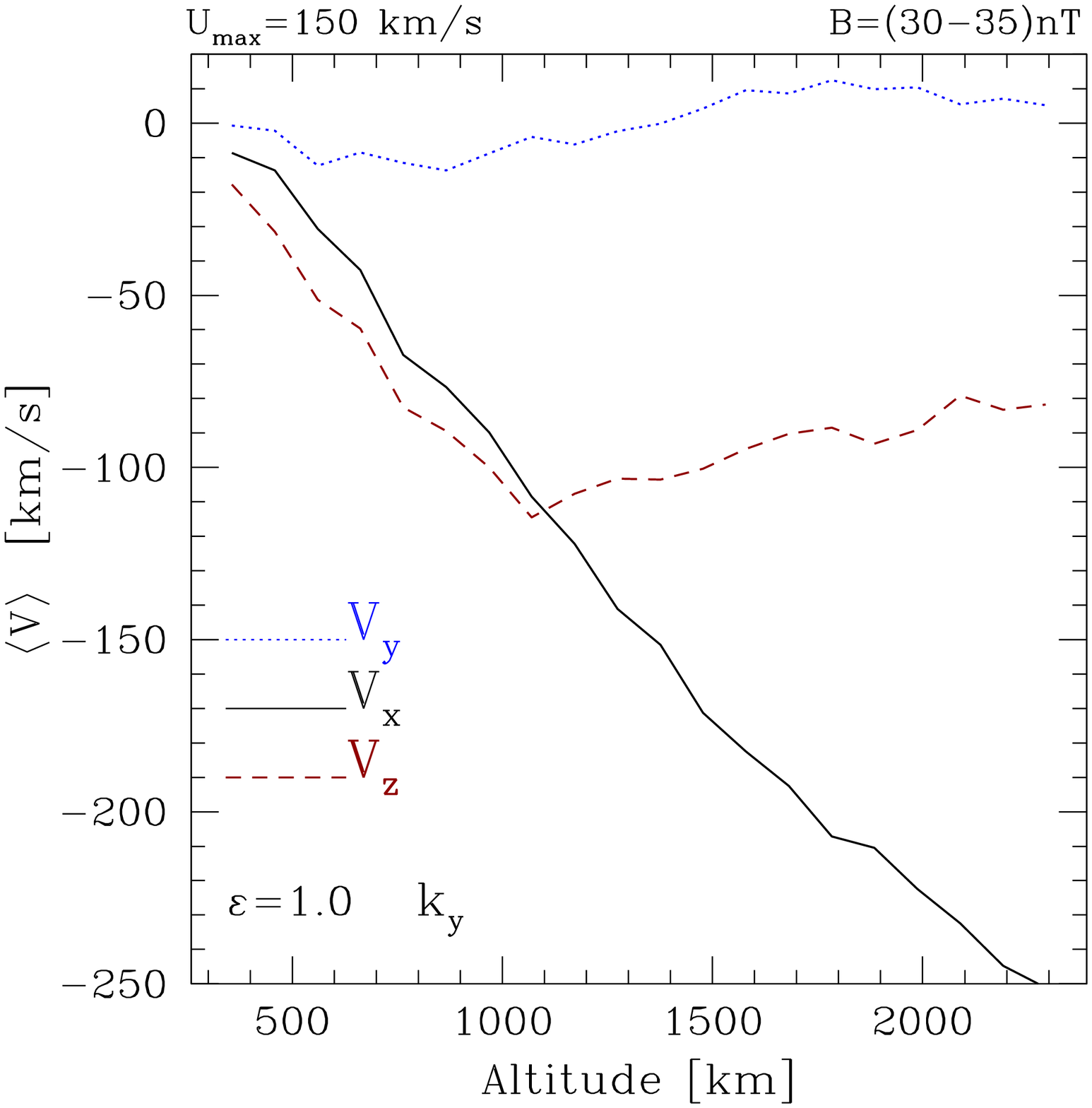}
\hspace{2pt}
\includegraphics[width=0.31\textwidth]{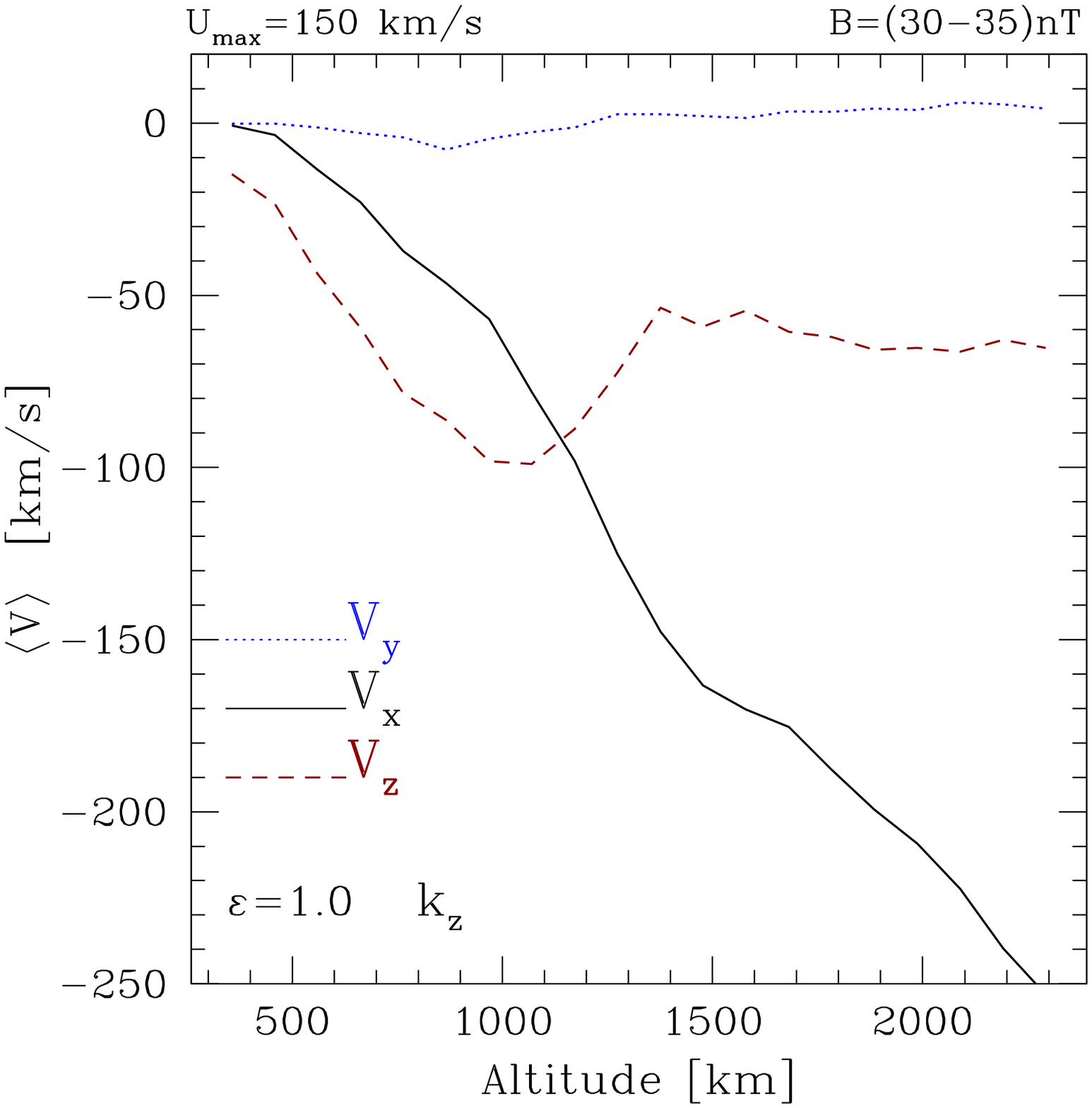}
\caption{As in Figure~\ref{fig:profiles} but with a stronger $B_{min}=30\,$nT and lower $U_{max}=150\,$km/s in order to favour an increase in the curvature of the $O^+$ pickup ions and its $V_x$ component. Note that the general behaviour of the profiles do not correspond to the ASPERA-3 measurements.}
\label{fig:profilesE}
\end{figure*}

In searching for a pure dynamical explanation of the velocity profiles measured by the Mars-Express spacecraft, we decided to use parameters in our model that would probably favor the dominance of the $V_x$ component over the $V_z$ component. For this, we would require a strong magnetic field at the poles, $B_{min}$,  and a small velocity $U_{max}$, since the gyro-radius of a particle, for the same $q/m$ ratio, is $r_g \propto v_\perp/B$ and this would make the trajectories bend more such that at the terminator the $X$--velocity would be increased. 

We decided to use for a new set of runs with $B_{min}=6 B_0=30\,$nT and $U_{max} = 0.375 V_{SW}=150\,$km/s, and keeping the rest of the parameters the same as in our fiducial case. As one might expect there is a dominance of the $V_x$ component over the $V_z$, but this occurs only for certain altitudes as shown in Figure~\ref{fig:phaseE} and \ref{fig:profilesE}. Even with these values of the parameters, rather extreme according to data (see, for example, Figure~\ref{fig:aspera} from where an estimate of $U_{max}$ can be obtained) the behaviour of the profiles do not correspond to the measurements; for example, no inversion of the velocity profiles at altitudes  $\approx 1000\,$km is present. This numerical experiment strongly suggests that even under more propense conditions, albeit of not being consistent with measurements of the properties of the $B$-field and stream velocity of the SW at the pole, the idea of the behaviour of $O^+$ being due to charge-particle dynamics is not sustainable.

 On the basis of our results, we conclude 
that turbulent-like magnetic field fluctuations, and the accompanying 
convective electric field variations, are not likely to influence the dynamics 
of picked-up $O^+$ ions in a sufficient manner to explain their essentially
tailward velocity measured over the magnetic pole of Mars. Furthermore, the relative behaviour of the different components of the velocity obtained in particle simulations does not correspond to what is measured at the Martian terminator. This   suggests that additional processes to a purely charged-particle dynamics must be playing a dominant role in the interaction of the collisionless solar wind and the ionospheric plasma in the region.

 A possibility is that advocated by Perez-de-Tejada and collaborators (e.g.~Perez-de-Tejada \& Dryer 1976, Perez-de-Tejada~1999, PdT09, RPAV10), where a viscous-like interaction of the SW plasma and the ionosheath plasma is occurring. 
The interaction of the SW with the upper ionosphere over
the magnetic polar regions of non-magnetic planets and objects is characterized by the existence of several features that lend themselves to be explained in terms of a fluid {\it viscous-like} interaction. 
Several of the phenomena for which it provides an explanation are briefly summarized here for the sake of completeness.
(1) The existence of a velocity shear in the ionosheath as spacecraft 
approach Venus over the magnetic poles (Romanov~et~al.~1979, Bridge~et~al.~1967, PdT09). The accompanying 
decrease in plasma density and the increase in the gas temperature are 
consistent with a viscous-like boundary layer in the region.
 (2) A similar situation is observed in the ionosheath of comet Halley below the so-called ``mystery transition'', where the decrease in SW velocity
and density and the simultaneous increase in the gas temperature, measured 
by the Giotto spacecraft, is consistent with a viscous-like flow 
characterized by an effective Reynolds number $\approx \! 30$ 
[Perez-de-Tejada 1989, RPAV10]. (3) 
 The transterminator flow observed by the 
Pioneer Venus Orbiter (PVO) is consistent with a viscous-like dragging 
of the Venus upper ionosphere by the SW flow (Perez-de-Tejada 1989). 
Since such interaction occurs preferentially over the magnetic polar regions 
this leads to the carving-out of plasma channels in the ionosphere and 
near wake, providing a simple explanation for the ionospheric holes 
measured by PVO in the nightside of the planet (P\'erez-de-Tejada 2001 \& 2004). It is important to indicate that  the physical nature of such ``anomalous viscosity'' is not addressed in these papers, and is the subject of current investigations.

\subsection{Final Remarks}

It must be stressed that the calculations presented in this paper are
not fully self-consistent, in the sense that the electromagnetic field and the solar wind velocity around the Martian ionosheath are assumed, not solved for. Nonetheless, the assumed SW velocity field and $B$-field structure are representative of what might be the Martian magnetic field in the region of interest; the case of the southern hemisphere would be more complex to model due to the presence of inhomogenieties due to crustal fields. 

The type of approach taken here clearly prevents us from studying  plasma instabilities which may play an important role in the coupling of solar wind and ionospheric plasma and significantly change our conclusions. The effect of such instabilities on the dynamics of $O^+$ ions, as well as that of magnetic field features related to its 3D character and the presence of magnetic concentration regions, are beyond the scope of the present study and will be treated in future contributions.

\acknowledgments
 
This research was funded by UNAM-PAPIIT Research Projects IN121406 and 
IN109710, and CONACyT Projects 25030 and 60354.



\end{document}